\documentclass[fleqn,usenatbib]{mnras}

\usepackage{newtxtext,newtxmath}
\usepackage[T1]{fontenc}

\DeclareRobustCommand{\VAN}[3]{#2}
\let\VANthebibliography\thebibliography
\def\thebibliography{\DeclareRobustCommand{\VAN}[3]{##3}\VANthebibliography}

\usepackage{graphicx}	
\usepackage{amsmath}	
\usepackage[version=4]{mhchem}
\usepackage{gensymb}
\usepackage{chemfig}

\let\oldAA\AA
\renewcommand{\AA}{\text{\normalfont\oldAA}}

\title[3D directly imaged cloud and chemistry connections]{\centering{Dynamically coupled kinetic chemistry in brown dwarf atmospheres - II. Cloud and chemistry connections in directly imaged sub-Jupiter exoplanets}}

\author[Lee, Tan and Tsai]{
Elspeth K.H. Lee$^{1}$, Xianyu Tan$^{2,3}$ and Shang-Min Tsai$^{4}$ \\
$^{1}$Center for Space and Habitability, University of Bern, Gesellschaftsstrasse 6, CH-3012 Bern, Switzerland \\
$^{2}$Tsung-Dao Lee Institute, Shanghai Jiao Tong University, 520 Shengrong Road, Shanghai, 200127, People’s Republic of China \\
$^{3}$School of Physics and Astronomy, Shanghai Jiao Tong University, 800 Dongchuan Road, Shanghai, 200240, People’s Republic of China \\
$^{4}$Department of Earth Sciences, University of California, 900 University Ave, Riverside CA 92521, California, US
}

\date{Accepted XXX. Received YYY; in original form ZZZ}

\pubyear{2023}

\begin{document}
\label{firstpage}
\pagerange{\pageref{firstpage}--\pageref{lastpage}}
\maketitle

\begin{abstract}
With JWST slated to gain high fidelity time dependent data on brown dwarf atmospheres, it is highly anticipated to do the same for directly imaged, sub-Jupiter exoplanets.
With this new capability, the need for a full 3D understanding to explain spectral features and their time dependence is becoming a vital aspect for consideration.
To examine the atmospheric properties of directly imaged sub-Jupiter exoplanets, we use the three dimensional Exo-FMS general circulation model (GCM) to simulate a metal enhanced generic young sub-Jupiter object.
We couple Exo-FMS to a kinetic chemistry scheme, a tracer based cloud formation scheme and a spectral radiative-transfer model to take into account the chemical and cloud feedback on the atmospheric thermochemical and dynamical properties.
Our results show a highly complex feedback between clouds and chemistry onto the 3D temperature structure of the atmosphere, bringing about latitudinal differences and inducing time-dependent stormy features at photospheric pressures.
This suggests a strong connection and feedback between the spatial cloud coverage and chemical composition of the atmosphere, with the temperature changes and dynamical motions induced by cloud opacity and triggered convection feedback driving chemical species behaviour.
In addition, we also produce synthetic latitude dependent and time dependent spectra of our model to investigate atmospheric variability and periodicity in commonly used photometric bands.
Overall, our efforts put the included physics in 3D simulations of exoplanets on par with contemporary 1D radiative-convective equilibrium modelling. 
\end{abstract}

\begin{keywords}
 planets and satellites: atmospheres -- stars: atmospheres -- planets and satellites: gaseous planets -- stars: brown dwarfs -- hydrodynamics -- methods: numerical 
\end{keywords}


\section{Introduction}

With the advent of JWST, the spectral properties of companion brown dwarfs and directly imaged planets will be observed with a large increase in precision, but also allow wide spectral coverage for the first time.
Several isolated brown dwarf atmospheric spectra have already been published using JWST data, for example, VHS 1256b \citep{Miles2023} and WISE 0359-54 \citep{Beiler2023}, allowing an order of magnitude greater understanding of their atmospheric processes and chemical compositions compared to previous instrumentation.
Companion brown dwarfs and exoplanets have also been directly imaged using the coronagraphic capabilities of JWST \citep[e.g.][]{Carter2023, Luhman2023}, revealing the spectral features across a wide wavelength range and providing an increasingly detailed characterisation of these enigmatic objects.
Several JWST Cycle 2 programs are expected to visit directly imaged companions (e.g. \#3181 P.I. Zhou, \#3514 P.I. Bonnefoy, \#3522 P.I. Ruffio, \#3647 P.I. Patapis), with some programs aiming to observe (or characterise) time-variability.
The increase in successful JWST proposals for these objects shows that directly imaged companions are quickly becoming an important part of the ensemble of exoplanets to consider for detailed modelling efforts.

Using simulated data and modelling \citet{Carter2021}, before the launch of JWST, suggested that the stellar-planet high-contrast capability of JWST would be able to observe sub-Jupiter exoplanets.
The JWST Early Research Program (ERS) \citep{Hinkley2022} directly imaged the super-Jupiter exoplanet HIP 65426b across the JWST coronagraphic modes of NIRCam ($\lambda$ = 2-5$\mu$m) and MIRI ($\lambda$ = 11-16$\mu$m). 
Their results presented in \citet{Carter2023} suggest that the flux contrast achieved by JWST is better than the pre-flight predictions from \citet{Carter2021}, and that future sub-Jupiter detection and characterisation is a distinct possibility.
\citet{Currie2023} review the state of ground-based efforts and the potential of JWST to directly image and observe exoplanets, in particular JWST's ability to observe past 10$\mu$m and confirm the spectral signature of silicate cloud particles in their atmospheres.
Evidence for such silicate absorption signatures is present in some isolated brown dwarf spectra from the Spitzer IRS instrument \citep[e.g.][]{Cushing2006,Suarez2022,Suarez2023} and in the recent JWST spectra of VHS 1256b \citep{Miles2023}.


Several observational campaigns have revealed the spectral and photometric variability of sub-stellar objects, providing evidence of time-dependent spatial inhomogeneity in their atmospheres.
\citet{Radigan2014} performed a ground-based survey of isolated brown dwarfs in the J, H and K$_{\rm s}$ bands, finding variability in 9 brown dwarf objects, with the strongest variable objects showing $>$2\% peak-to-peak relative flux differences occurring at the L-T transition.
Several studies observed spectral variability in brown dwarfs using HST WFC3 \citep{Buenzli2012,Apai2013,Yang2015,Lew2016,Lew2020}. 
These studies use binned spectra-photometry bands of WFC3 data, finding differences in the variability strength and phases inside and out the \ce{H2O} absorption band.
These studies suggest that the variability is driven by a thinning and thickening of the cloud deck on these brown dwarfs rather than strong spatial patchiness. 
\citet{Zhou2022} observe VHS 1256b with HST WFC3, finding a large $\approx$33\% peak-to-peak variability in its spectral flux, suggesting that strongly changing inhomogenous weather patterns are present in VHS 1256b's atmosphere.
\citet{Tannock2021} observe variability on timescales of $\approx$1 hour of three fast rotating brown dwarfs, which suggests that changes in weather patterns on these fast rotators occur quicker than a rotational period.
The analysis performed in \citet{Vos2017,Vos2018,Vos2020} and \citet{Suarez2023b} suggests that the broadband variability strength of the brown dwarfs depends strongly on the viewing angle of the observations, showing that the equatorial regions of brown dwarfs show stronger spectral indications of variability compared to the polar regions. 
\citet{Apai2021} used TESS data to measure the periodic variability of the Luhman 16 AB system across almost 22 days of observations, calculating a rotation rate for both brown dwarfs and a long term periodic component of $\sim$90.8 hours for the system.
\citet{Vos2022} specifically look at a set of young sub-stellar objects, taking spectra and Spitzer measurements. 
They also find measurable variability occurring in these young objects, which are good analogues to directly imaged exoplanet objects.
\citet{Biller2017} and \citet{Artigau2018} provide reviews of the numerous studies and observational campaigns looking for variability in brown dwarf atmospheres.
In addition to these variability studies, \citet{Crossfield2014} mapped the atmosphere of the closest brown dwarf Luhman 16AB using the Doppler imaging technique, finding that the atmosphere was highly inhomogenous and changed with the rotation of the brown dwarf, suggesting atmospheric patchiness is primarily responsible for the variability signatures in these atmospheres.

Several modelling studies have looked at processes that can explain the spectral variability in brown dwarf and directly imaged exoplanet atmospheres.
In 1D studies, \citet{Robinson2014} used a time-dependent radiative-convective model with periodic thermal perturbations to explore the timescales of spectral and thermal responses in the atmosphere.
They find that convectively driven thermal perturbations can be a key part of explaining the spectral variability of brown dwarf atmosphere, but deep perturbations take time to propagate to the upper photospheric pressures.
\citet{Morley2014b} model patchy sulfide and salt cloud layers and atmospheric thermal perturbations to simulate variability of T and Y dwarfs.
\citet{Tan2019} use a coupled tracer cloud model with diffusion to model time-variability of vertical cloud structures in brown dwarf atmospheres.
\citet{Tremblin2020} apply modified temperate-pressure (T-p) profiles to mimic thermal variability and also investigate the effects of non-equilibrium chemistry on the variability and strength of spectral features.
\citet{Luna2021} use the \citet{Ackerman2001} equilibrium cloud model to produce vertical cloud profiles  that attempt to fit the Spitzer IRS silicate feature in brown dwarfs \citep{Cushing2006,Looper2008} and Spitzer photometric band variability \citep{Metchev2015}.

In 3D studies, \citet{Showman2013} used the MITgcm with an idealised anelastic system of equations, suitable for the deep brown dwarf atmospheric regions.
They find that temperature fluctuations of a few Kelvin can be initialised by the large scale flow.
\citet{Zhang2014} used a global shallow-water model with a pulsing thermal perturbation scheme, which imitates temperature changes induced from convective motions, to investigate the possible dynamical regimes of brown dwarfs, they find a large difference in the peak-to-peak variability percentage depending on the particular dynamical regime of the brown dwarf atmosphere.
\citet{Tan2021b} use GCM simulations of brown dwarfs to explore the effect of clouds on the variability and dynamical properties of the atmosphere, showing that the viewing latitude affects the observed periodicity of the variability as well as the rotation rate of the brown dwarf.
\citet{Tan2022} coupled the tracer model of \citet{Tan2021b} to brown dwarf regime GCMs, performing simulations across a wide range of rotation rates and other atmospheric parameters.
They find that the cloud coverage results in periodic, latitude-dependent modulation of the outgoing flux, showing how the coupled atmospheric dynamics and clouds at the viewing angle of the planet affect their observed variability metrics.
\citet{Lee2023a} coupled the mini-chem kinetic chemistry scheme to a GCM and performed several 3D tests of the non-equilibrium scheme across the L, T and Y dwarf effective temperature regime.
They find that storms and global scale dynamical features can produce localised enhancements or depletion of chemical species.
\citet{Hammond2023} use thermally perturbed shallow water models to explore the dynamical regimes of brown dwarf atmospheres, finding four distinct dynamical regimes depending on the thermal Rossby number and radiative timescale of the dynamical layer.

In addition to the above modelling studies, \citet{Apai2013,Apai2017} and \citet{Apai2021} developed a data driven spot and band system to fit to variability curve data.
Their model attempts to reconstruct the possible large scale atmospheric features on the brown dwarf or exoplanet from the normalised flux density periodicity.

Overall, the outlook for characterising the time variability of directly imaged exoplanets in the near future is highly promising and warrants detailed theoretical and modelling efforts.
A common conclusion to the above studies was the need to explore the coupled problem of clouds, chemistry and thermal perturbations which feedback in a self-consistent manner, which our current study attempts to address.
In particular, the effect of dis-equilibrium chemistry has important effects on the observed spectra of these objects \citep[e.g.][]{Saumon2000,Miles2020,Miles2023}, which is a key addition to our model.

In this study, we perform a three dimensional general circulation model (GCM) simulation of a generic metal enhanced young sub-Jupiter object.
We include the mini-chem kinetic chemistry scheme, a tracer based equilibrium cloud formation scheme and spectral radiative-transfer scheme coupled into the Exo-FMS GCM.
In Section \ref{sec:improv}, we detail the additions and improvements to the model compared to our previous studies.
In Section \ref{sec:GCM}, we present the directly imaged exoplanet GCM set-up and results of the GCM simulation.
In Section \ref{sec:PP}, we present post-processing results of the GCM output and produce synthetic spectra along with time dependent photometric band variability maps.
Section \ref{sec:disc} presents a discussion of our results and Section \ref{sec:conc} the conclusion of our study.

\section{Additions to previous model}
\label{sec:improv}

In \citet{Lee2023b} (hereafter LTT) we used a simplified GCM simulation to explore the 3D chemical properties of brown dwarfs by coupling a kinetic chemistry model, `mini-chem' \citep{Tsai2022, Lee2023a}, to the Exo-FMS GCM \citep[e.g.][]{Lee2021}, with the atmospheric dynamics driven by stochastic vertical temperature perturbations of the atmosphere through convective motions \citep{Showman2013, Tan2022}.
Mini-chem is a miniature chemical kinetics network model that uses net-forward reaction rate tables to greatly reduce the number of reactions and species evolved in the system \citep{Tsai2022}.
The N-C-H-O network of mini-chem contains 12 species with 10 reactions, 6 of which are net-forward reaction tables, making it highly efficient when coupled to large scale 3D atmosphere models.
However, this efficiency comes at the cost of some accuracy in predicting the mixing ratio of species when compared to the full kinetic model \citep{Tsai2022}.

Briefly, the main simplifications used in LTT were
\begin{itemize}
    \item Rosseland mean grey radiative-transfer scheme.
    \item Cloud free atmospheres.
    \item Simple tracer convection with dry convective adjustment.
\end{itemize}
In this current study, we remove these limitations and perform more self-consistent modelling of these types of atmospheres.
We add several physical processes important for a fuller and realistic simulation of these atmospheres, namely, 
\begin{itemize}
    \item Mixing length theory.
    \item Tracer convection through vertical diffusion.
    \item Tracer based equilibrium cloud model.
    \item Spectral radiative-transfer.
\end{itemize}
which we detail below.
The convective thermal perturbation scheme, where we assume a storm timescale of $\tau_{\rm storm}$ = 10$^{5}$ s, and mini-chem kinetic chemistry are the same as in LTT.

\subsection{Mixing length theory}
\label{sec:MLT}

Due to the global scales of the GCM simulation, convective cell motions which occur on length scales of kilometers or less, much smaller than the GCM spatial resolution, are not able to be resolved. 
Therefore, a parametrisation scheme must be used to capture the net large scale effects of convection onto the atmosphere.
For this, we use mixing length theory (MLT) inside the GCM to model the changes in temperature due to convective motions in the deep atmosphere and upper secondary convective zones.
We use the simple mixing length formulation outlined in \citet{Joyce2023}, overall following a similar approach to that described in \citet{Robinson2014, Marley2015} and \citet{Tan2019}, except using vertical pressure units more natural to the GCM vertical grid.
For completeness, we summarise the pressure based equation set used in the GCM below.

The convective region is defined when the vertical atmospheric temperature gradient is larger than the adiabatic gradient ($\nabla$ $>$  $\nabla_{\rm ad}$), 
with the local atmospheric temperature gradient given by
\begin{equation}
    \nabla = \frac{\partial\ln T}{\partial\ln p},
\end{equation}
where T [K] is the gas temperature and p [pa] the gas pressure.
The adiabatic gradient is given by
\begin{equation}
    \nabla_{\rm ad} = \frac{R_{\rm d}}{c_{\rm p}} = \kappa_{\rm ad},
\end{equation}
where R$_{\rm d}$ [J kg$^{-1}$ K$^{-1}$] is the specific gas constant of the air, c$_{\rm p}$ [J kg$^{-1}$ K$^{-1}$] the specific heat capacity of the air and $\kappa_{\rm ad}$ the adiabatic coefficient of the air.
The goal of mixing length theory is to calculate the rate of convergence to the adiabatic temperature lapse rate from the current difference between the temperature and adiabatic lapse rates.
This is performed by assuming that rising localised parcels of air carry a thermal flux across a `mixing length' of atmosphere, which then tends the atmosphere to the adiabatic T-p profile.
This is different to dry convective adjustment, where typically the T-p profile is adjusted time independently towards the adiabatic profile directly through enthalpy conservation.

The vertical convective heat flux in convective regions, F$_{\rm conv}$ [J m$^{-2}$ s$^{-1}$], derived from mixing length theory is given by \citep[e.g.][]{Joyce2023}
\begin{equation}
    F_{\rm conv} = \frac{1}{2}\rho c_{\rm p} w T \frac{L}{H_{\rm p}} \left(\nabla - \nabla_{\rm ad}\right),
\end{equation}
where $\rho$ [kg m$^{-3}$] is the mass density of the air and $w$ [m s$^{-1}$] the characteristic vertical velocity and $L$ [m] the characteristic mixing length.
From this equation we can see that when the local temperature gradient equals the adiabatic gradient there is static stability and no convective flux is present.

The characteristic mixing length, $L$, is typically assumed to be some factor $\alpha$ of the atmospheric pressure scale height, H$_{\rm p}$ [m], 
\begin{equation}
\label{eq:mixl}
    L = \alpha H_{\rm p} = \alpha \frac{R_{\rm d}T}{g},
\end{equation}
where g [m s$^{-2}$] is the gravity.
\citet{Robinson2014} state that their 1D  brown dwarf models were not too sensitive to factors of $\alpha$ between 1/2 and 2, so we assume an $\alpha$ = 1 value in our scheme.
We note however, much discussion in the literature over the past century has been dedicated to examining MLT and the mixing length parameter, from the original \citet{Bohm-Vitense1958} derivation to finessing for optically thin and thick convection zones \citep[e.g.][]{Cox1968,Mihalas1978} to full-spectrum turbulence theory \citep[e.g.][]{Canuto1991}.
\citet{Joyce2023} provide an overview of these approaches and consequences of the $\alpha$ parameter on astrophysical problems.

The characteristic vertical velocity can be given with buoyancy arguments through the local Brunt-V{\"a}is{\"a}l{\"a} frequency \citep[e.g.][]{Marley2015}
\begin{equation}
\label{eq:wconv}
   w = L \left[\frac{g}{H_{\rm p}}\left(\nabla - \nabla_{\rm ad}\right)\right]^{1/2},
\end{equation}
where the relation \citep{Joyce2023}
\begin{equation}
   \frac{\partial T}{\partial z} = - \frac{T}{H_{\rm p}}\frac{\partial\ln T}{\partial\ln p},
\end{equation}
is used to alter the equivalent expression in \citet{Marley2015}.
The temperature tendency, $\partial T$/$\partial t$ [K s$^{-1}$], of an atmospheric layer due the convective heat flux is then
\begin{equation}
    \left(\frac{\partial T}{\partial t}\right)_{\rm conv} = \frac{g}{c_{\rm p}}\frac{\partial F_{\rm conv}}{\partial p}.
\end{equation}
In the GCM, F$_{\rm conv}$ can be calculated at the mid-point layers and linearly interpolated to the level edges to calculate the net convective flux into each convective layer.

We extensively tested the MLT scheme with our 1D radiative-convection time stepping routine that emulates directly the radiative-transfer (RT) and convective approach in the GCM. 
As in \citet{Robinson2014}, we find a smaller timestep ($\lesssim$ 1 s) is required for numerical stability when evolving the temperature using MLT due to the usually large convective fluxes.
We therefore use a sub-timestep approach, where the MLT scheme is integrated in smaller timestep increments until the GCM dynamical timestep is reached.
The net temperature tendency of a layer can then be passed back to the GCM for regular integration.
In testing, we found near identical results occur between MLT and the dry convective adjustment scheme over a handful of integration steps, suggesting that the strong convective fluxes in MLT extremely efficiently evolve the T-p gradient to the adiabat.

\subsection{Tracer vertical diffusive mixing}
\label{sec:diff}

Using mixing length theory, a thermal eddy diffusion coefficient in convective zones, K$_{\rm zz}$ [m$^{2}$ s$^{-1}$], is given by the relation \citep[e.g.][]{Marley2015}
\begin{equation}
    K_{\rm zz} = wL,
\end{equation}
where w is the characteristic vertical velocity from Eq. \eqref{eq:wconv} and L the characteristic mixing length from Eq. \eqref{eq:mixl}.
In this way an eddy diffusion coefficient can be derived in each convective region of the atmosphere, which is not otherwise possible when using a convective adjustment scheme \citep[e.g.][]{Marley2015}.
In the GCM, this results in a 3D localised, spatially dependent K$_{\rm zz}$ value for each cell.
This is different to the globally derived K$_{\rm zz}$ value, typically assumed or derived from GCMs or theoretical efforts \citep[e.g.][]{Parmentier2013,Steinrueck2021, Tan2022}.
Convection resolving models can also provide a direct estimate of K$_{\rm zz}$ \citep[e.g.][]{Lefevre2022}.

Some momentum carried vertically from convective motions will pass through the radiative-convective boundary (RCB) into the radiative region, perturbing the atmosphere near the radiative-convective boundary.
This `convective overshoot' effect has been noted in several 2D and 3D mesoscale convection resolving models of M dwarf and brown dwarf atmospheres \citep[e.g.][]{Freytag2010,Lefevre2022}.
To account for this in 1D models, \citet{Woitke2004, Woitke2020} propose an exponential fitting function dropoff for the turbulent vertical velocity beyond the RCB through
\begin{equation}
    \ln w = \ln w_{\rm rcb} - \beta \cdot \rm{max}\left(0, \ln p_{\rm rcb} - \ln p\right),
\end{equation}
where $\beta$ is a parameter between 0 and $\approx$2.2 derived from fitting the \citet{Freytag2010} mesoscale hydrodynamic models.
We use this parameterisation to estimate K$_{\rm zz}$ above the radiative-convective boundary in the GCM, assuming $\beta$ = 2.2,
with a small minimum K$_{\rm zz}$ = 10$^{1}$ m$^{2}$ s$^{-1}$ applied to stabilise the scheme.

With the calculation of K$_{\rm zz}$, we can now apply a diffusion scheme to transport chemical tracers in the vertical direction to emulate the mixing rates from convective motions.
To perform the vertical diffusion of tracers inside the GCM, we follow a first order explicit time-stepping method.
We note that tracers are still advected in the horizontal and vertical direction through the bulk dynamical flow in addition to the vertical diffusive component.

\subsection{Cloud formation model}

To explore the effects of clouds and their feedback onto the atmospheric chemical and temperature structure, we use a tracer based simplified equilibrium cloud formation model based on \citet{Tan2021, Tan2021b} and \citet{Komacek2022}, with some differences to better fit our current GCM setup.
In this system, a tracer representing the condensable vapour volume mixing ratio, q$_{\rm v}$, is evolved alongside a tracer representing the condensed vapour volume mixing fraction, q$_{\rm c}$.
The two main evolution equations are given as
\begin{equation}
\label{eq:qv}
    \frac{dq_{\rm v}}{dt} = (1 - s) \frac{min(q_{\rm s} - q_{\rm v},q_{\rm c})}{\tau_{\rm c}} - s \frac{(q_{\rm v} - q_{\rm s})}{\tau_{\rm c}},
\end{equation}
and
\begin{equation}
\label{eq:qc}
    \frac{dq_{\rm c}}{dt} = s \frac{(q_{\rm v} - q_{\rm s})}{\tau_{\rm c}} - (1 - s) \frac{min(q_{\rm s} - q_{\rm v},q_{\rm c})}{\tau_{\rm c}},
\end{equation}
where $\tau_{\rm c}$ [s] is the condensation timescale and s the thermal stability coefficient, taking a value of 0 or 1. 
Here we take the stability coefficient to be a function of the supersaturation, S, of the species
\begin{equation}
    S = \frac{q_{\rm v} p}{p_{\rm vap}(T)},
\end{equation}
where p$_{\rm vap}$(T) [Pa] is the temperature dependent saturation vapour pressure of the chosen species. 
Where S $>$ 1, s = 1 and S $<$ 1, s = 0. 
If S = 1, there is thermal stability and the rate of change of the vapour and condensate tracers are set to zero.
The supersaturation ratio also defines the equilibrium tracer volume mixing ratio, q$_{\rm s}$, in Eqs. \eqref{eq:qv} and \eqref{eq:qc} through taking S = 1, the equilibrium saturation value
\begin{equation} 
    q_{\rm s} = min \left(\frac{p_{\rm vap}}{p}, 1\right).
\end{equation}
In this study, we assume a \ce{MgSiO3} cloud composition, taking the vapour pressure expression from \citet{Ackerman2001}.
We assume a deep vapour volume mixing ratio of q$_{\rm v}$ = 10$^{-4}$, approximately 10 times the atomic Mg gas phase Solar fraction at chemical equilibrium in the deep atmosphere \citep{Woitke2018}.
The vapour and cloud tracer initial conditions are assumed to be zero, after which condensable vapour is transported from the lower boundary at its deep value through advection and convection the same way as the chemical tracers.

In \citet{Komacek2022} the vertical transport of the condensable vapour tracer was performed using an additional vertical mixing timescale term, mimicking the replenishment of condensable vapour to the lower layers by convective transport from the (unmodeled) interior, after which the vapor tracers are transported to the upper layers by vertical advection and convective adjustment.
This returned the lowest layer vapour mixing ratio to a deep value, q$_{\rm deep}$, over a prescribed mixing timescale, $\tau_{\rm deep}$ [s].
Here, we instead couple the vapour tracer to the vertical diffusive mixing scheme (Sect. \ref{sec:diff}) to trace the convective mixing time-dependently in-line with the other gas phase chemical tracers, and assume a constant deep lower boundary value for q$_{\rm v}$.  
The condensate fraction is also coupled to the diffusion scheme, which is an assumption that the cloud particles are well coupled to the small scale eddies with a low Stokes number \citep[St $\ll$ 1, e.g.][]{Woitke2020}.

Our method is essentially a tracer based equilibrium cloud condensation model with a relaxation timescale.
This allows greater flexibility and ease when coupled to the GCM compared to coupling a microphysical model such as mini-cloud \citep{Lee2023c}, however, cloud size distribution properties are required to be parameterised in the equilibrium scheme.
We follow a log-normal distribution for ease of computation, prescribed by a median (or geometric mean) particle size and variance.

For the log-normal distribution, the total number density, N$_{0}$ [m$^{-3}$], of the cloud particle size distribution is given by
\begin{equation}
\label{eq:N0}
    N_{0} = \frac{3\epsilon q_{\rm c}\rho}{4\pi \rho_{\rm c}r_{\rm m}^{3}}\exp\left(-\frac{9}{2}\ln^{2}\sigma\right),
\end{equation}
where $\epsilon$ is the molecular weight fraction of the condensate composition to the background atmosphere, $\rho$ [kg m$^{-3}$] is the local atmospheric gas density, $\rho_{\rm c}$ [kg m$^{-3}$] the bulk density of the condensed material, r$_{\rm m}$ [m] the median particle radius and $\sigma$ the geometric standard deviation of the distribution.
The area weighted or effective radius, r$_{\rm eff}$ [m], of the distribution is given by
\begin{equation}
\label{eq:reff}
    r_{\rm eff} = r_{\rm m} \exp\left(\frac{5}{2}\ln^{2}\sigma\right).
\end{equation}
Assuming a uniform and constant bulk density for each cloud particle size, the mass weighted or volume weighted mean radius, r$_{\rm v}$ [m], is given by
\begin{equation}
\label{eq:rV}
    r_{\rm v} = r_{\rm m} \exp\left(\frac{7}{2}\ln^{2}\sigma\right).
\end{equation}
We note that r$_{\rm v}$ is equivalent to the \citet{Ackerman2001} (Eq. 13) characteristic settling radius, r$_{\rm w}$, when f$_{\rm sed}$ = 1 and $\alpha$ = 1.
Typically, the \citet{Ackerman2001} method assumes $\alpha$ > 1 to ensure r$_{\rm w}$ > r$_{\rm v}$.

Using the results of previous equilibrium and microphysical cloud model studies, we can now chose suitable parameters for the log-normal size distribution.
For $\sigma$, a value $\sim$1.5-2 is usual, which covers a wide size-distribution where coagulation processes are important in setting the width of the distribution, for example, the \citet{Ackerman2001} framework typically assumes $\sigma$ = 2.
However, microphysical studies using bin-based models \citep[e.g.][]{Powell2018, Gao2018b} for exoplanet clouds suggest bi- or multi- modal distributions are more physical, with small and large particle modes present, suggesting a single continuous distribution does not capture the full properties of the cloud particles sufficiently.
\citet{Gao2018} attempt to fit log-normal distributions in the \citet{Ackerman2001} framework to CARMA microphysical model results for KCl clouds in Y-dwarf atmospheres, where they find the log-normal fitting can decently capture the large particle mode but fails to capture the small-particle mode.
Therefore, following the results and conclusions in \citet{Gao2018}, we use a slightly narrower log-normal distribution with $\sigma$ = 1.5 than typically used in equilibrium cloud modelling \citep[e.g.][]{Ackerman2001} and only try to capture the large particle mode which provides the majority of the cloud opacity.

Defining a median particle size is more difficult, with microphysical studies showing it to be sensitive to the vertical mixing rates, gravity (through settling rates) and the details of the thermochemical environment.
The microphysical model results of \citet{Powell2018} for the hot Jupiter regime, and \citet{Gao2018} for the Y-dwarf regime, suggest an effective particle size of $\sim$1$\mu$m is appropriate for the mixing and gravity parameters of our simulation.
However, the exact effective particle size to assume is uncertain for our specific T-dwarf temperature structure and metal enhanced regime. 
Therefore, we treat the effective radius as a parameter, and assume r$_{\rm eff}$ = 1$\mu$m.
As a result, the log-normal distribution has a median and volume weighted mean particle size of $\approx$0.66$\mu$m and $\approx$1.18$\mu$m respectively.

For the condensable tracer, we perform the vertical transport due to particle settling the same way as in \citet{Lee2023c}, using an explicit second order McCormack integrator with minmod flux limiter.
We calculate the vertical settling velocity the same way as in \citet{Lee2023c}, which takes into account the local atmospheric dynamical viscosity using the square root mixing law for a mixed composition gas from \citet{Rosner2012}.
We use the volume weighted mean radius of the distribution as the representative settling velocity particle size.

\subsection{Spectral radiative-transfer}

In LTT, a simple Rosseland mean grey opacity scheme was applied using the fitting function from \citet{Freedman2014}. 
This scheme was not able to capture the radiative feedback from the changing volume mixing ratio (VMR) of chemical species resultant from the kinetic chemistry scheme.
Here we use an updated spectral model, able to provide this chemical feedback in a consistent manner, as well as include the effect of cloud opacity feedback on the temperature structure.
For the longwave radiative-transfer model we use the \citet{Toon1989} method which includes multiple scattering, commonly used in hot exoplanet GCM simulations \citep[e.g.][]{Showman2009,Roman2017}.

\subsubsection{Gas phase opacity}

For the gas phase opacities we use a correlated-k scheme with 11 bands the same as \citet{Kataria2013}, which was found to be suitable for hot exoplanet modelling.
During the staged spin-up time, we use a pre-mixed correlated-k table assuming CE at 10x Solar \citep{Asplund2021} metallicity.
When coupled to mini-chem, opacities are computed on the fly and k-tables mixed using the adaptive equivalent extinction method presented in \citet{Amundsen2017}.
We include the line opacity of all molecular species included in mini-chem, that is OH \citep{Hargreaves2019}, \ce{H2O} \citep{Polyansky2018}, CO \citep{Li2015}, \ce{CO2} \citep{Yurchenko2020}, \ce{CH4} \citep{Hargreaves2020}, \ce{C2H2} \citep{Chubb2020}, \ce{NH3} \citep{Coles2019}, HCN \citep{Harris2006}, where the citation for each species denotes the source of the line-list used to produce the correlated-k opacity tables.
We include collisional induced absorption of \ce{H2}-\ce{H2}, \ce{H2}-He, \ce{H2}-H and He-H, with data taken from the HITRAN database \citep{Karman2019}.
For Rayleigh scattering, we include \ce{H2}, He and H.

The VMR of each species is input into the opacity mixer directly from the GCM and mini-chem results for that timestep.
Our scheme allows the changing VMR of each species to alter the local opacity and feed back into the heating/cooling of the atmosphere, a notable improvement to LTT Rosseland mean scheme.

The opacity of Na and K is well known to be important in setting the observational properties of T-dwarfs at optical wavelengths \citep[e.g.][]{Burrows2003} and affecting the T-p structure \citep[e.g.][]{Malik2019}.
We include the opacity of Na and K into the opacity scheme, however, the chemical kinetic details of Na and K are not well known in the brown dwarf and exoplanet atmosphere context.
To estimate the abundances of Na and K, we assume that both Na and K are quenched in a similar manner to other chemical species to their deep chemical equilibrium abundances, this leads to an approximate constant VMR of 10$^{-5}$ for Na and 10$^{-6}$ for K at 10x Solar metallicity \citep[e.g.][]{Woitke2018}.
The Kurucz line-list database is used to calculate the opacity of Na and K \citep{Kurucz1995}.
In addition, we assume that the strong optical wavelength absorbers present in L-dwarf atmospheres (for example TiO, VO, FeH, Fe) are fully rained out and do not contribute their opacity to the atmosphere.

\subsubsection{Cloud particle opacity}

In our cloud scheme, the extinction efficiency, single scattering albedo and asymmetry parameter of a range of cloud particle sizes can be pre-calculated assuming a normalised log-normal size distribution for each wavelength bin of the correlated-k scheme.
The extinction opacity can then be calculated using the total number density of the condensed phase following Eq. \eqref{eq:N0} and the log-normal distribution parameters.
The distribution can then be integrated to find the size-distribution integrated opacity, single scattering albedo and asymmetry parameter.
This avoids expensive on-the-fly Mie theory calculations such as those performed in \citet{Lee2023c}.
For the pre-calculated tables, we use the LX-MIE code from \citet{Kitzmann2018}, and assume amorphous \ce{MgSiO3}, with the optical constants also taken from \citet{Kitzmann2018}.
As in \citet{Lee2016, Lee2023c}, a boxcar smoothing of the cloud opacity is performed in the vertical direction to avoid large opacity gradients at the cloud free and cloudy boundary which leads to numerically unstable heating rates from the RT scheme.

\section{Directly imaged sub-Jupiter model}
\label{sec:GCM}

\begin{table}
\centering
\caption{Adopted Exo-FMS GCM simulation parameters for the directly imaged sub-Jupiter atmosphere simulation. We use a cubed-sphere resolution of C96 ($\approx$ 384 $\times$ 192 in longitude $\times$ latitude).}
\begin{tabular}{c c c l}  \hline \hline
 Symbol & Value  & Unit & Description \\ \hline
 T$_{\rm int}$ & 1000 & K & Internal temperature \\
 P$_{\rm 0}$ & 100 &  bar & Reference surface pressure \\
 P$_{\rm up}$ & 10$^{-4}$ &  bar & Upper boundary pressure \\
 c$_{\rm p}$ & 11183  &  J K$^{-1}$ kg$^{-1}$ & Specific heat capacity \\
 R$_{\rm d}$ & 3220 &  J K$^{-1}$ kg$^{-1}$  & Specific gas constant \\
 $\kappa$ &  0.288 & -  & Adiabatic coefficient \\
 g$_{\rm p}$ & 10  & m s$^{-2}$ & Acceleration from gravity \\
 R$_{\rm p}$ & 7.149 $\cdot$ 10$^{7}$  & m & Planetary radius\\
 $\Omega_{\rm p}$ & 1.745 $\cdot$ 10$^{-4}$ & rad s$^{-1}$ & Rotation rate \\
 M/H & 1 & - & $\log_{10}$ solar metallicity \\
 P$_{\rm amp}$ & 10 & bar & Perturbation boundary layer \\
 T$_{\rm amp}$ & 3.71 $\cdot$ 10$^{-4}$ & K s$^{-1}$ & Perturbation amplitude \\
 $\alpha$ & 1 & - & MLT scale parameter \\
 $\beta$ & 2.2 & - & Overshooting parameter \\
 $\tau_{\rm storm}$ & 10$^{5}$ & s & Storm timescale \\
 $\tau_{\rm drag}$ & 10$^{6}$ & s & Basal drag timescale \\
 $\Delta$ t$_{\rm hyd}$ & 60  & s & Hydrodynamic time-step \\
 $\Delta$ t$_{\rm rad}$  & 3600 & s & Radiative time-step \\
 $\Delta$ t$_{\rm MLT}$ & 0.5  & s & Mixing length theory time-step \\ 
 $\Delta$ t$_{\rm ch}$  & 3600 & s & Mini-chem time-step \\
 $\tau_{\rm c}$ & 120  & s & Eq. cloud timescale \\
 N$_{\rm v}$ & 60  & - & Vertical resolution \\
 d$_{\rm 4}$ & 0.16  & - & $\mathcal{O}$(4) divergence dampening coef. \\
\hline
\end{tabular}
\label{tab:GCM_parameters}
\end{table}

In this section, we present the results of our young sub-Jupiter simulation.
The parameters of the GCM simulation are presented in Table \ref{tab:GCM_parameters}.
We assume a metallicity of 10x Solar, with a Jupiter-like rotation rate of 10 hours.
This captures a `middle ground' dynamical regime described in LTT.
As in LTT a deep `basal drag' is included, implemented as a Rayleigh drag term \citep[e.g.][]{Tan2021}, with a surface drag of $\tau_{\rm drag}$ = 10$^{6}$ s at the 100 bar lower boundary which is linearly decreased in strength to a pressure of 50 bar.
Kinetic energy dissipated by this drag is returned to the atmospheric energy budget through a localised increase in temperature.
This drag timescale allows large scale features to develop without being damped out.
\citet{Tan2022} explore in depth the effects of changing this deep drag value on the global atmospheric dynamical properties.

We follow a staged approach to spin-up towards the fully coupled model. 
For the first stage, we use the Rosseland mean opacity scheme from LTT without clouds and chemistry but with MLT to spin-up the model for 2000 days, which is approximately double the simulation time for statistical equilibrium to be achieved in this regime \citep{Tan2022}.
For the second stage, we switch to the pre-mixed correlated-k scheme for 100 days, including the diffusion and settling of cloud vapour and particles but without the opacity component.
We then slowly ramp up the cloud opacity for the next 100 days to account for the cloud opacity feedback, which is then retained for the rest of the simulation.
For the third and final stage, mini-chem and the opacity mixing scheme is switched on and allowed to evolve time dependently with the flow and diffusive scheme.
We assume CE for the initial conditions of the chemical tracers to ensure a smooth transition between equilibrium and non-equilibrium chemical abundance feedback onto the RT scheme.
This full scheme is run until convergence of the kinetic-chemistry scheme, which occurs at a global scale at around a total simulation time of 2980 days.
The results presented here are snapshots of the output at this time.

\begin{figure*}
    \centering
    \includegraphics[width=0.49\textwidth]{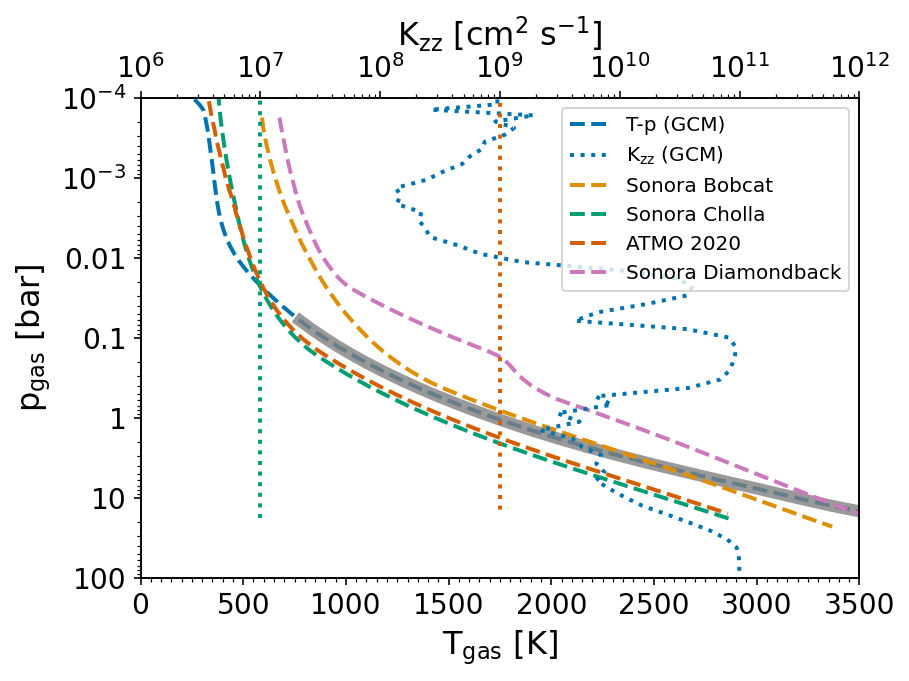}
    \includegraphics[width=0.49\textwidth]{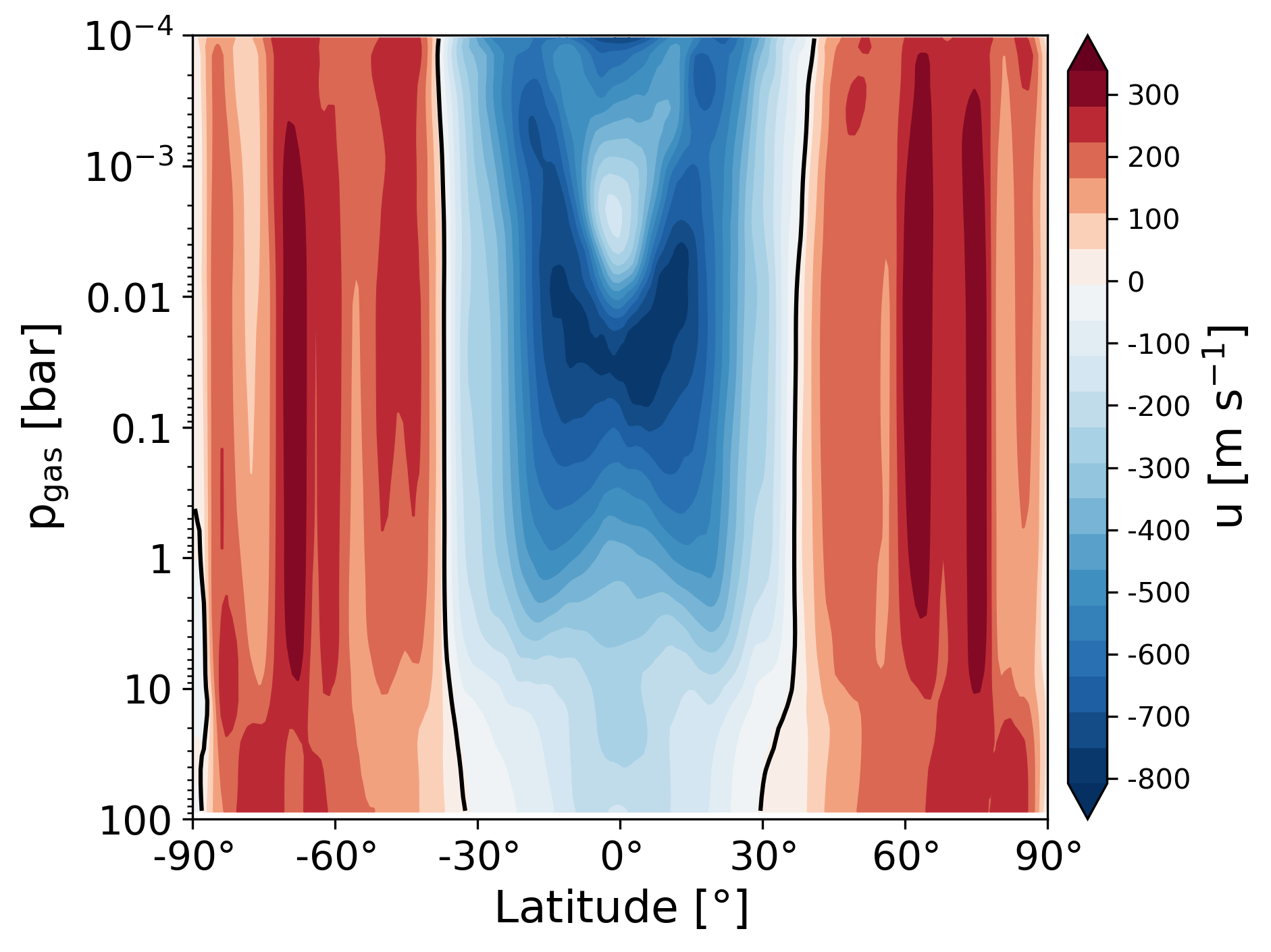}
    \includegraphics[width=0.49\textwidth]{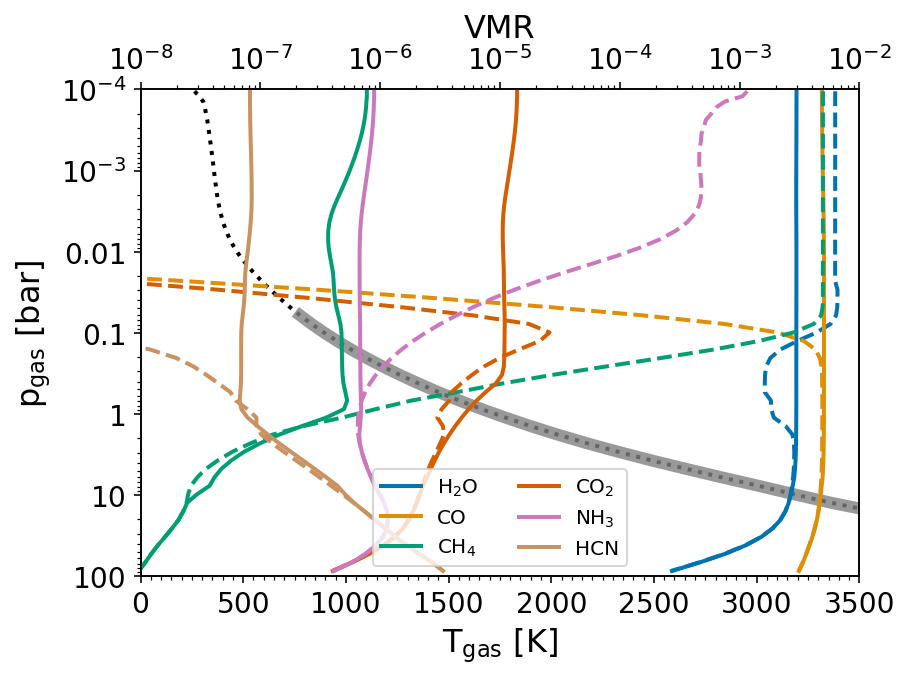}
     \includegraphics[width=0.49\textwidth]{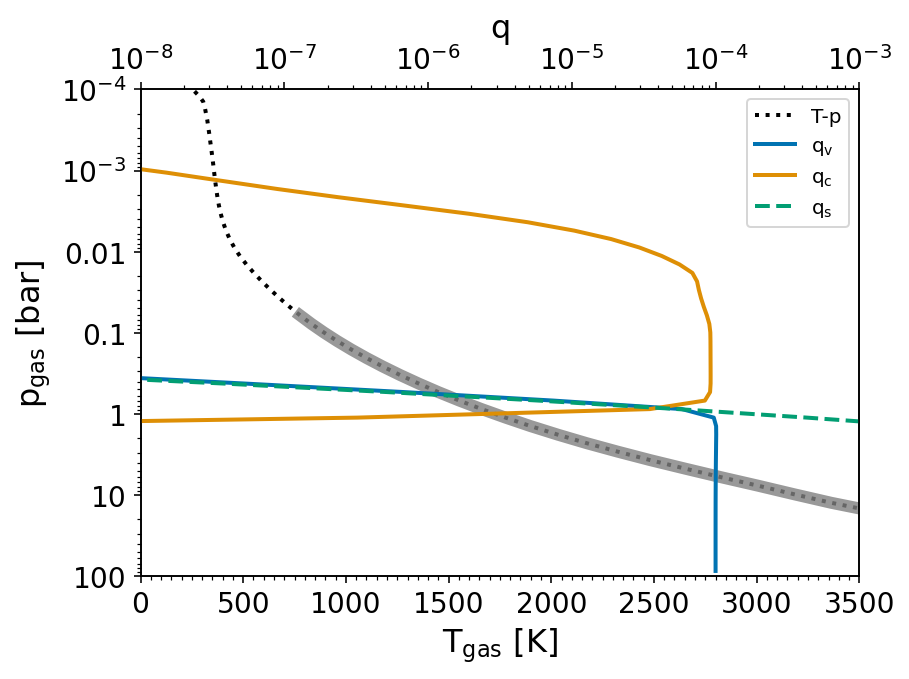}
    \caption{End of simulation snapshot results from the young sub-Jupiter GCM simulation.
    Top left: global averaged K$_{\rm zz}$ vertical profile.
    Top right: zonal mean zonal velocity.
    Lower left: global averaged VMR of select chemical species from mini-chem (solid lines) and assuming chemical equilibrium (dashed lines).
    Lower right: global averaged cloud vapour fraction q$_{\rm v}$, condensate fraction q$_{\rm c}$ and saturation vapour fraction q$_{\rm s}$.
    The global averaged T-p profile is given by the black dotted line, with the convective region denoted by the grey shaded region.}
    \label{fig:GCM_1D}
\end{figure*}

\begin{figure}
    \centering
    \includegraphics[width=0.49\textwidth]{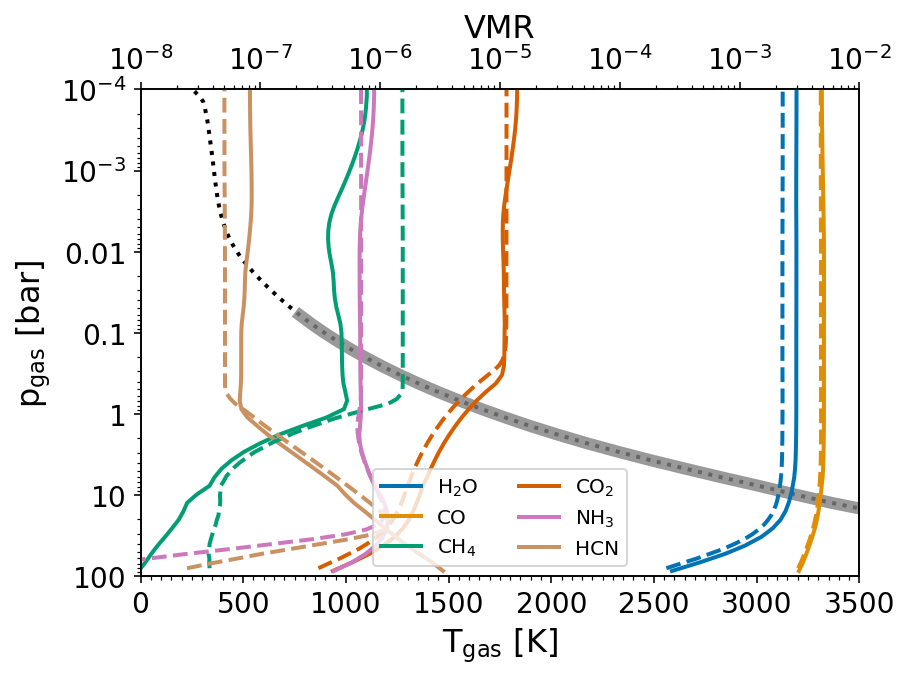}
    \caption{Comparison between the end of simulation snapshot global averaged mini-chem results from the GCM (solid lines) for each chemical species and the 1D VULCAN C-H-O-N network model (dashed lines) which used the GCM derived global averaged T-p and K$_{\rm zz}$ values as input.}
    \label{fig:VULCAN}
\end{figure}

\begin{figure*}
    \centering
    \includegraphics[width=0.49\textwidth]{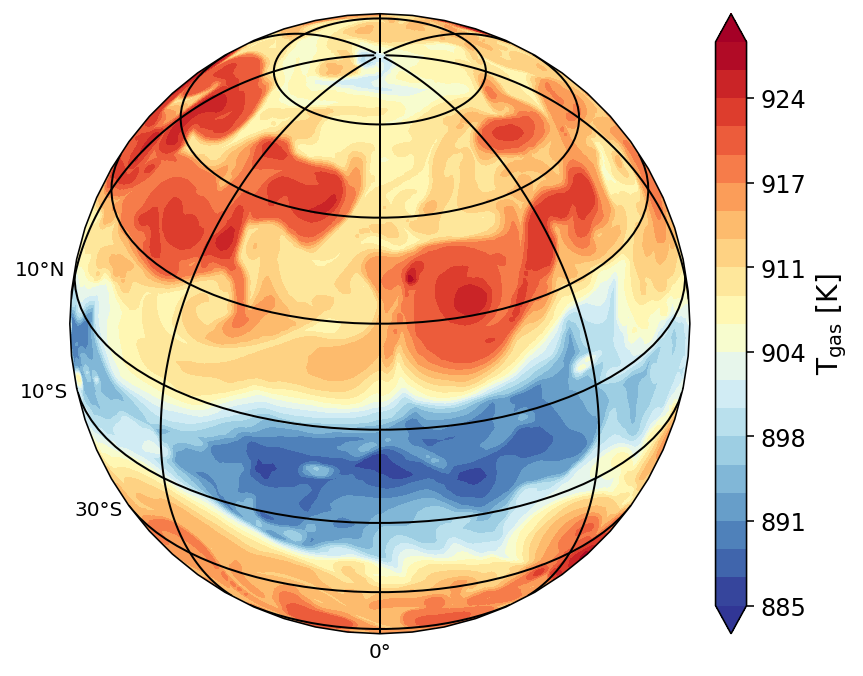}
    \includegraphics[width=0.49\textwidth]{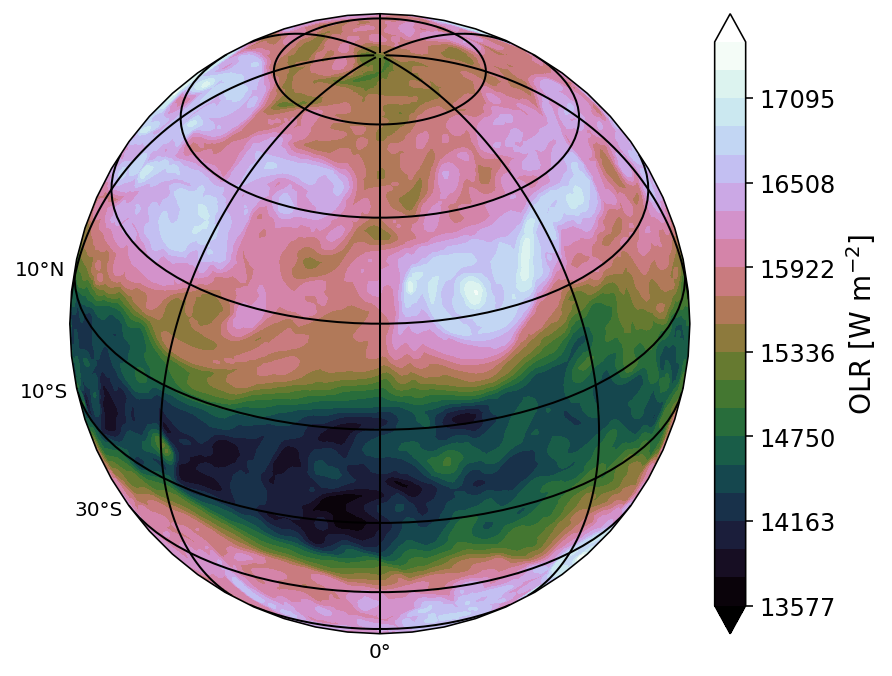}
    \includegraphics[width=0.49\textwidth]{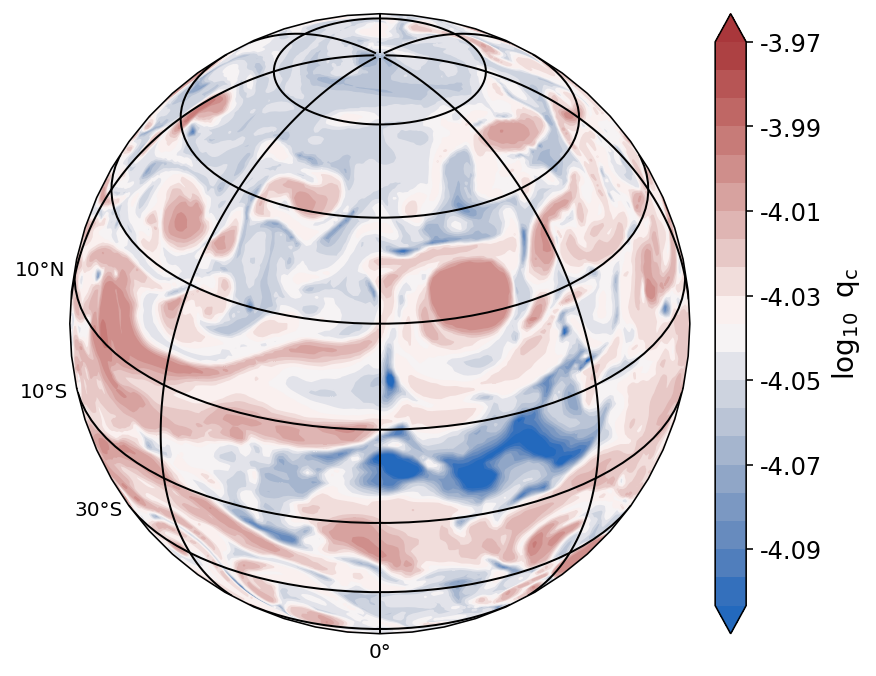}
    \includegraphics[width=0.49\textwidth]{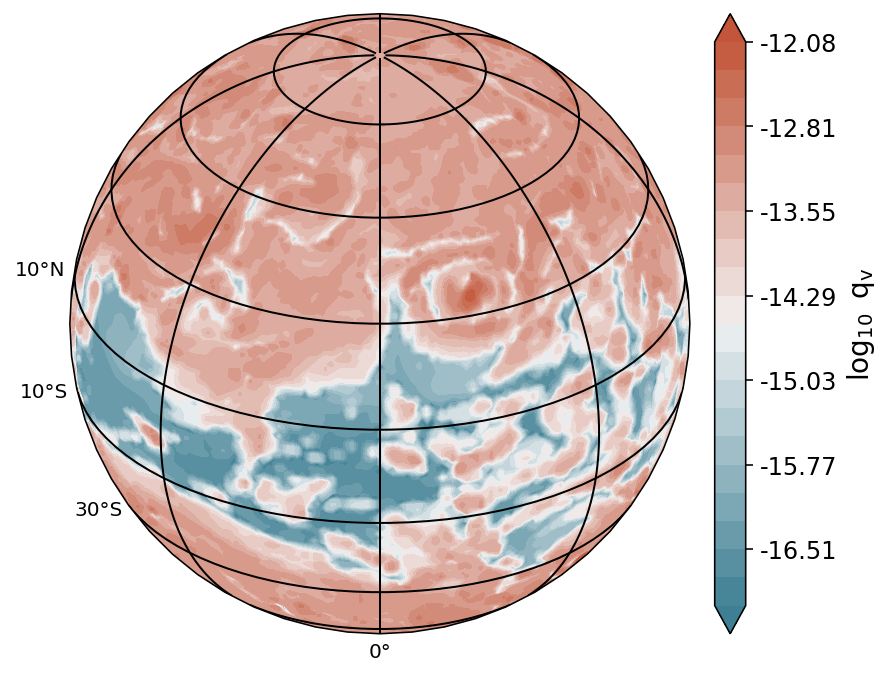}
    \caption{End of simulation (at 2980 days) snapshot results from the young sub-Jupiter GCM simulations, spatial distribution projections at the 0.1 bar pressure level, near the photosphere of the planet.
    Top left: atmospheric temperature.
    Top right: outgoing longwave radiation.
    Bottom left: condensate mass fraction.
    Bottom right: condensate vapour fraction.}
    \label{fig:GCM_proj}
\end{figure*}

\begin{figure*}
    \centering
    \includegraphics[width=0.49\textwidth]{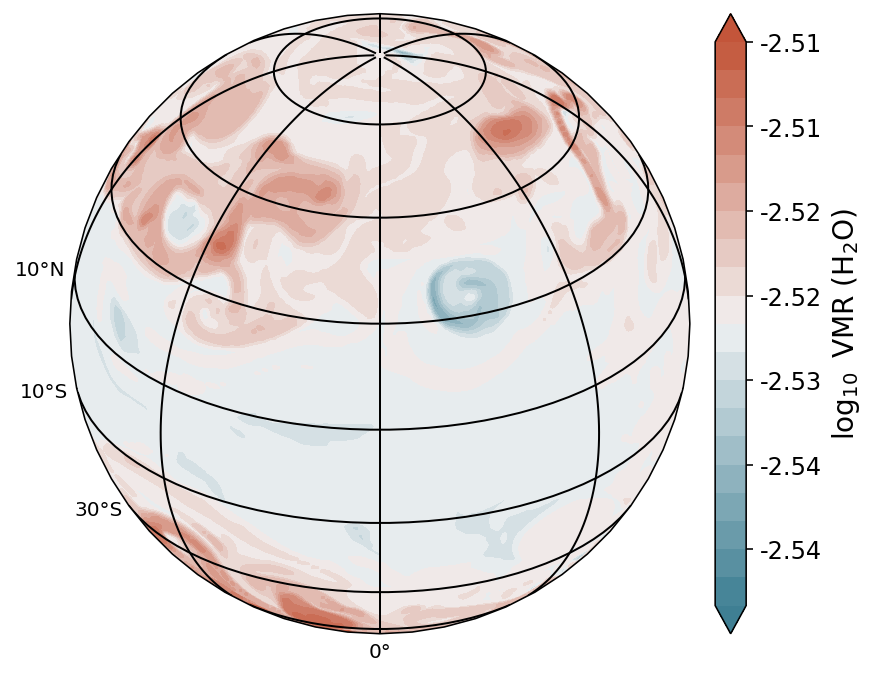}    
    \includegraphics[width=0.49\textwidth]{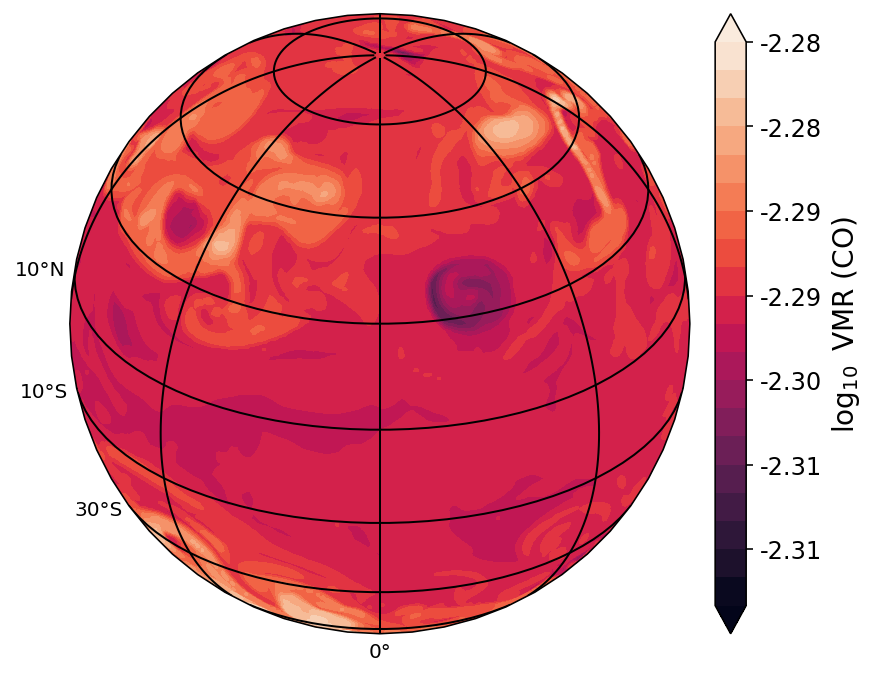}
    \includegraphics[width=0.49\textwidth]{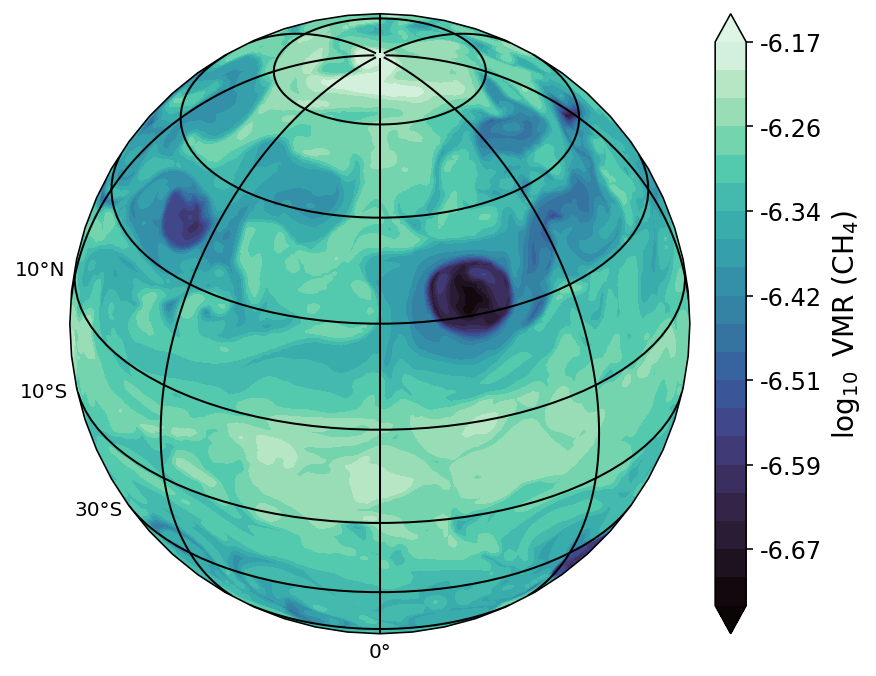}
    \includegraphics[width=0.49\textwidth]{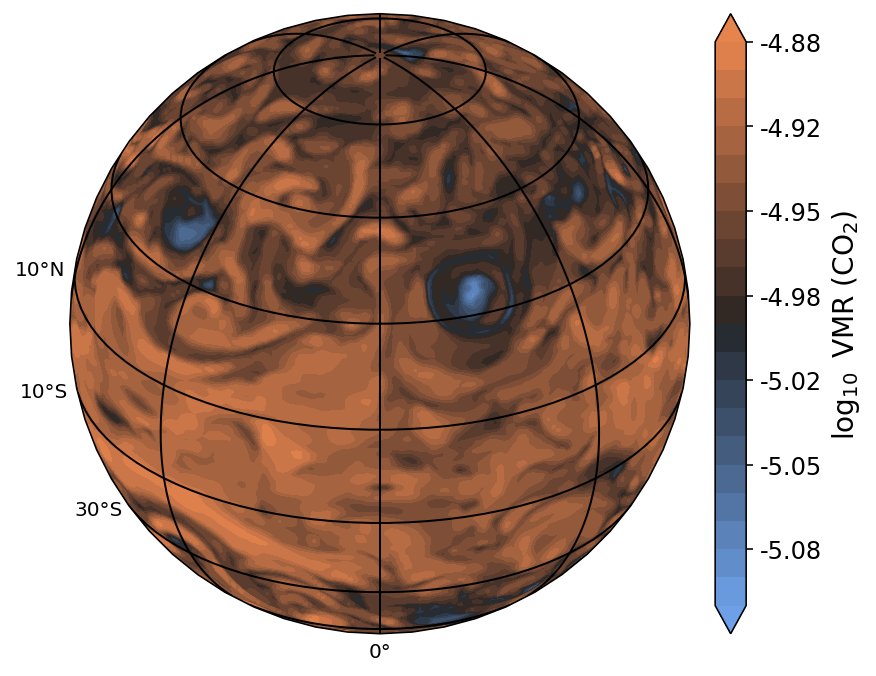}    
    \includegraphics[width=0.49\textwidth]{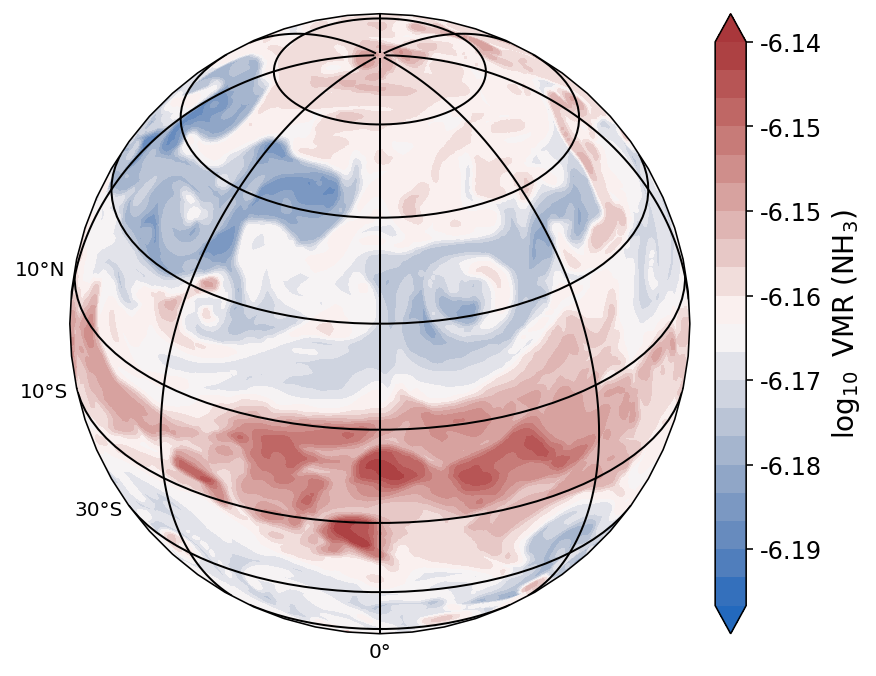}
    \includegraphics[width=0.49\textwidth]{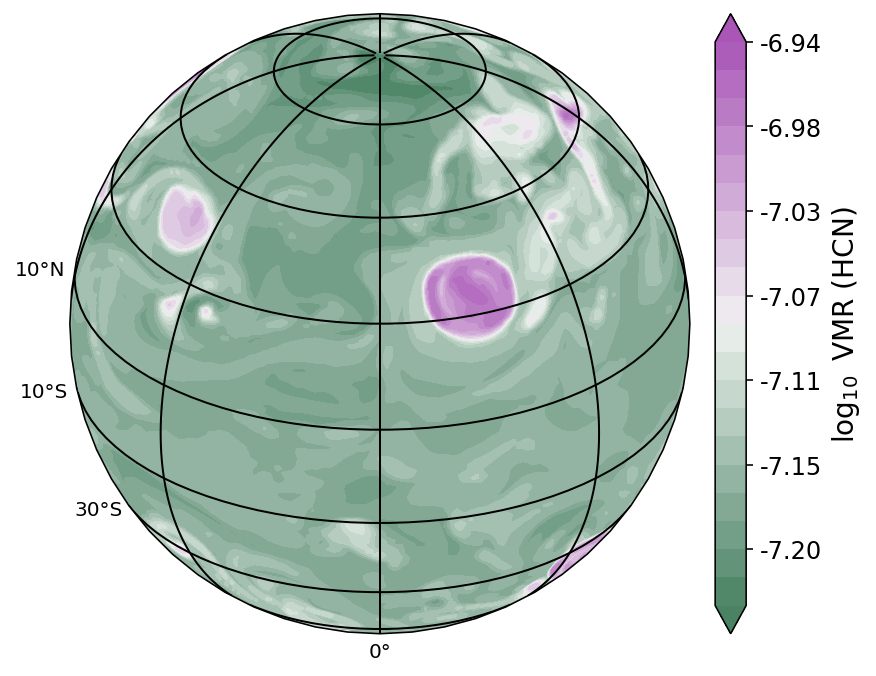}        
    \caption{End of simulation (at 2980 days) snapshot results from the young sub-Jupiter GCM simulation spatial distribution projections of the kinetic chemistry results at the 0.1 bar pressure level, near the photosphere of the planet.
    Top left: volume mixing ratio of \ce{H2O}.
    Top right: volume mixing ratio of \ce{CO}.
    Middle left: volume mixing ratio of \ce{CH4}.
    Middle right: volume mixing ratio of \ce{CO2}.
    Bottom left: volume mixing ratio of \ce{NH3}.
    Bottom right: volume mixing ratio of HCN.}
    \label{fig:GCM_proj2}
\end{figure*}

\subsection{Global averaged results}

In Figure \ref{fig:GCM_1D}, we present 1D globally averaged values derived from the GCM simulation.
To provide context for our results, we compare the averaged GCM T-p profile to T$_{\rm eff}$ = 1000 K, $\log$ g = 3, cloud free 1D RCE models from the literature; 1x Solar metallicity non-equilibrium chemistry model from \citet{Karalidi2021} (Sonora Cholla) and \citet{Phillips2020} (ATMO 2020), and the $\approx$3x Solar metallicity chemical equilibrium Sonora Bobcat model \citep{Marley2021}.
We also compare to the T$_{\rm eff}$ = 1000 K, $\log$ g = 3.5, $\approx$3x Solar, f$_{\rm sed}$ = 1 Sonora Diamondback T-p profile (C. Morley priv. comm.).
The assumed constant K$_{\rm zz}$ values from \citet{Karalidi2021} and \citet{Phillips2020} are also over plotted to compare to the GCM derived K$_{\rm zz}$ value.

The global averaged GCM T-p profile compares well with the 1D radiative-convective-equilibrium (RCE) models, generally being hotter than the 1x Solar non-equilibrium models, but cooler than the Sonora Bobcat model.
The impact of clouds on the T-p profile is only slight, with a small bend up in temperature between the 1 and 0.01 bar pressure levels, where the cloud particles reside.
Such a feature is common to 1D RCE and 3D convection resolving brown dwarf models that include a cloud component \citep[e.g.][]{Burrows2006,Witte2009,Marley2010,Lefevre2022} and arises from a combination of the additional absorption of infrared radiation from the cloud opacity, but also the longwave scattering greenhouse effect which reduces the cooling effectiveness of the atmosphere \citep[e.g.][]{Mohandas2018}.

The zonal mean zonal velocity plot in Figure \ref{fig:GCM_1D}, shows a typical velocity pattern for this regime \citep{Tan2022}, with a strong counter rotating broad equatorial jet structure and higher latitude co-rotating bands.
A notable slowdown in the jet speed occurs at around 10$^{-3}$ bar, which is also seen for simulations in a similar parameter regime \citep{Tan2021b}.

Figure \ref{fig:GCM_1D} also presents the globally averaged chemical profiles from the simulation, compared to the chemical equilibrium (CE) values.
Here it is clear the atmosphere is highly out of chemical equilibrium, with all species showing signs of vertical quenching behaviour.
Taking into account this chemical non-equilibrium behaviour into the radiative-transfer scheme in a consistent manner is known to be important in shaping the T-p profiles of hot exoplanets \citep[e.g.][]{Drummond2016}.
All species quench around the 1 bar pressure level, except CO, which departs from the equilibrium value at around the 0.1 bar level.
Of note is the strong departure from equilibrium of \ce{CH4} and CO which have sharp transitions from their CE abundances.

The global averaged cloud vapour fraction and cloud condensate fraction from Figure \ref{fig:GCM_1D} show a typical strongly mixed equilibrium cloud formation structure \citep[e.g.][]{Ackerman2001, Rooney2022}, where the vapour fraction remains constant up to the condensation level and then converted into cloud condensate above this level. 
The balance between the upward mixing and settling of cloud particles give rise to the tapering off structure with height \citep[e.g.][]{Gao2018b}, with the bulk of the cloudy region in our simulation occurring between 1 and 0.01 bar.
Due the strong mixing in the cloud particle region, the condensed mass fraction remains relatively constant inside the cloud region.
The vertical size of the cloud deck and the strong K$_{\rm zz}$ region coincide between $\sim$1 and 0.01 bar, this suggests that the cloud opacity perturbations on the temperature induce a strong localised convective response in the atmosphere.
A similar vertical mixing response was noted in the \citet{Tan2019} study, which produced similar K$_{\rm zz}$ profile shapes.
A small virga region is present below the main cloud deck, where settling clouds take time to evaporate beyond their thermal instability temperature.

\subsection{Comparison to VULCAN}

In this section, we use the global averaged T-p and K$_{\rm zz}$ profile from the GCM as input to the 1D kinetic-chemistry model VULCAN \citep{Tsai2017,Tsai2021}. 
We then compare the resultant chemical profiles from VULCAN to those produced by the GCM simulation.
Figure \ref{fig:VULCAN} presents this comparison.
Photochemistry is not activated in the VULCAN simulations.
The VULCAN results line up reasonably well with the GCM global averaged results, with the notable exception of \ce{CH4} which quenches at slightly higher pressure in the GCM compared to VULCAN.
In addition, the deep atmosphere abundances for \ce{CH4}, HCN and \ce{NH3} are quite different between the models.
Most likely responsible for these differences is, as noted in \citet{Tsai2022}, that the species abundance between mini-chem and VULCAN can be in error of up to 100$\%$ or more, depending on the thermochemical and mixing profiles.
However, the reduced and fast chemical scheme of mini-chem allows computationally efficient coupling to GCMs which is its main trade-off benefit.
The GCM also contains an extra vertical transport term through vertical advection, not included in the diffusion only VULCAN comparison. 
This may push the quench level to deeper in the atmosphere in the GCM if vertical motions are vigorous enough. 
In addition, we have used a globally averaged K$_{\rm zz}$ for the VULCAN and ignored local variance in K$_{\rm zz}$ which may also overall increase mixing rates in the GCM.

However, due to the overall similarity between the VULCAN and GCM results, this exercise shows a highly useful way to more quickly converge future models of this type.
We suggest that future modelling efforts use a large constant K$_{\rm zz}$ for the chemical tracers in the GCM to quickly converge the chemical species, then later relax to the K$_{\rm zz}$ given by MLT.
An alternative option would be to try use 1D kinetic models such as VULCAN, or the 1D RCE models that include non-equilibrium chemistry as an initial condition for the chemical tracers in the GCM, however, from our experience in this study, this might lead to thermal stability issues if the opacity structure is radically different to that given by chemical equilibrium or other given T-p initial condition.

\subsection{Global atmospheric features}

In Figure \ref{fig:GCM_proj}, we present orthographic projections of the GCM output, namely the temperature, outgoing longwave radiation (OLR), the cloud condensate mixing ratio, q$_{\rm c}$, and the cloud vapour mixing ratio, q$_{\rm v}$.
Most apparent in the results is that the cloud condensate spatial distribution has regions of correlation with temperature, but also large regions with little or no correlation.

The cloud condensate fraction follows a highly inhomogeneous pattern across the globe, with stronger patches of cloud appearing at both cooler (equatorial) and warmer regions (mid-latitudinal storm regions) of the planet. 
There is a reduction in cloud coverage at the high latitude and polar regions, while the mid-latitude storm and equatorial regions show distinct cloud bands, storms and general patchiness of stronger and weaker cloud coverage.
These mid-latitude cloud bands follow the temperature pattern well, but do not exactly correlate to the temperature structure.
An equatorial cloud band is present, but does not correlate to the temperatures structure there, these clouds follow a more dynamically driven pattern rather than temperature pattern.

Overall, our cloudy model results suggest that the clouds are inducing a significant feedback onto the temperature and dynamical profile of the planet in a inhomogenous and localised manner.
The dynamical range of the cloud mass fraction value varies between $\approx$10$^{-3.97}$ and $\approx$10$^{-4.09}$ at the 0.1 bar pressure level, which is around a 25\% variation.
Our results and analysis of the simulation suggest that for the high- and mid-latitude regions, perturbations in the cloud coverage impact the local temperature structure and induce warming storms. 
This warming as well as dynamical motions dissipate the cloud over time, periodically warming and cooling the atmosphere due to this cloud feedback.
At the equatorial regions, our results show that the cloud coverage is highly dynamically driven and more variable with larger dynamical range of the cloud fraction, likely a result of larger scale equatorial waves triggered by cloud feedback \citep{Tan2021b}.
Our results show a strong and complex coupling between the large scale dynamical properties of the atmosphere in this dynamical regime, the formation of storms and the feedback of clouds onto this system.
Similar conclusions were found in coupled cloud models in \citet{Tan2021b} and \citet{Tan2022} using a more simple GCM set-up.

In Figure \ref{fig:GCM_proj2} we present projections of the chemical results of the GCM simluation at 0.1 bar.
In comparing these results to those of Figure \ref{fig:GCM_proj}, there is a distinct correlation between the chemical composition and stormy regions of the atmosphere.
Most apparent is that the mid-latitude spatial distribution of species is highly correlated to the cloud coverage. 
In the center of the stormy cloudy regions there is a strong increase of HCN and decrease in \ce{CH4}, with the other species decreasing more slightly in VMR.
However, just outside the cloudy regions, at similar temperature to the cloudy centers, the VMR can sharply return to the background mean or increase depending on the species.
This suggests that the chemical species respond dynamically to the formation of storm regions, quickly forming chemically distinct storm regions as the storm regions appear.
Then after the storm dissipates the chemical species relax back to the background mean over time, mostly following the resultant temperature perturbation of the atmosphere due to the storm.

\subsection{Properties of storm regions}

\begin{figure*}
    \centering
    \includegraphics[width=0.49\textwidth]{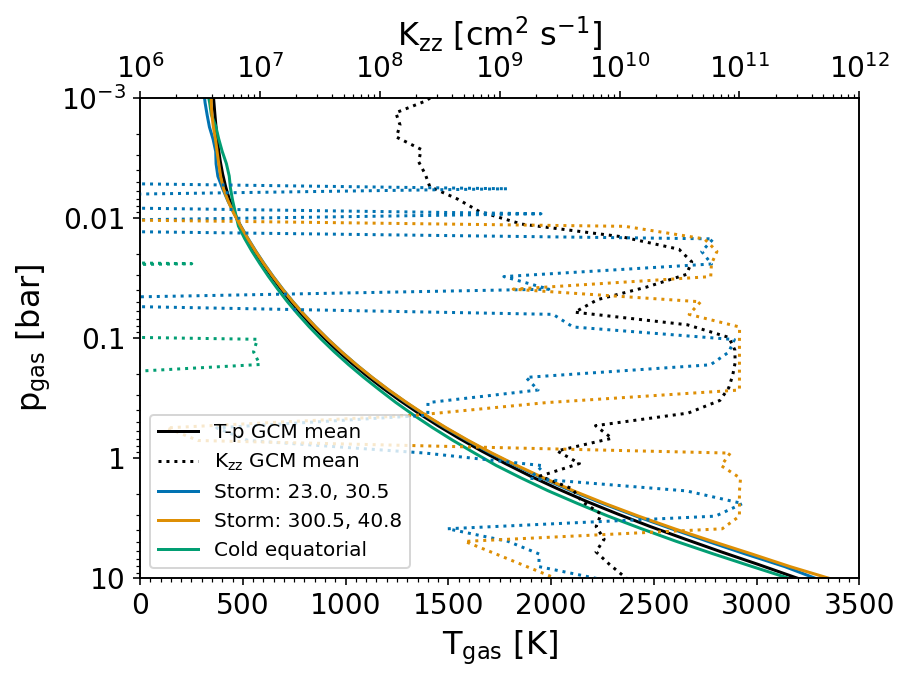}
    \includegraphics[width=0.49\textwidth]{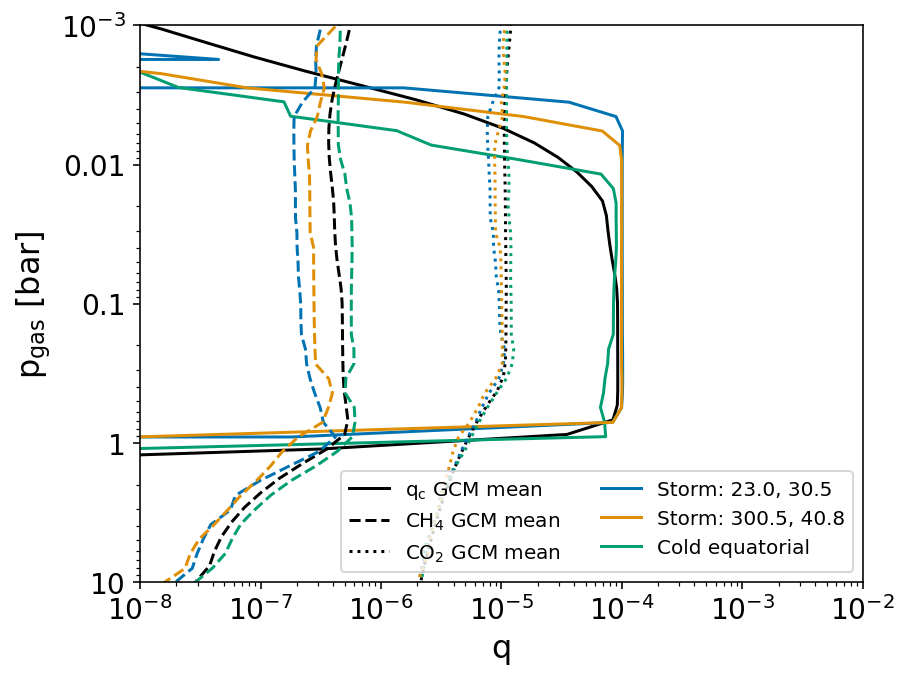}
    \caption{Left: T-p (solid lines) and K$_{\rm zz}$ (dotted lines) profiles of the global mean (black), the storms at longitude, latitude (23.0, 30.5, blue) and (300.5, 40.8, orange) and a cold equatorial profile (green).
    Right: condensation fraction q$_{\rm c}$ (solid), CH$_{4}$ VMR (dashed) and CO$_{2}$ VMR (dotted) global mean (black), the storms at longitude, latitude (23.0, 30.5, blue) and (300.5, 40.8, orange) and a cold equatorial profile (green).}
    \label{fig:inout}
\end{figure*}

\begin{figure}
    \centering
    \includegraphics[width=0.49\textwidth]{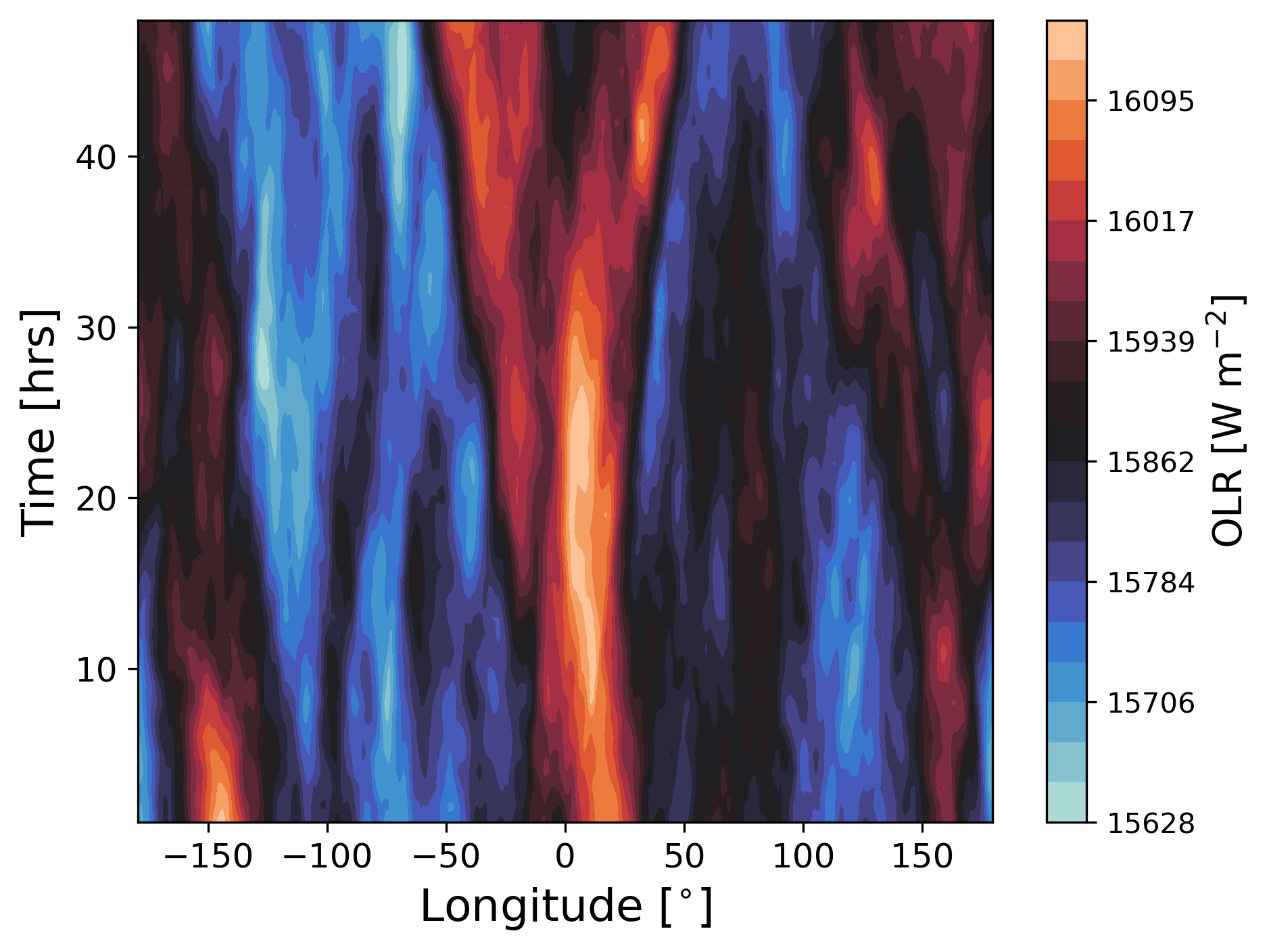}
    \caption{Hovm{\"o}ller diagram averaged across all latitudes for 42 hours of simulation after the end point. 
    This shows a general decrease in the OLR across the globe past hour 30, where the atmosphere enters a more quiescent phase.}
    \label{fig:GCM_ST}
\end{figure}

In Figure \ref{fig:inout}, we show T-p, K$_{\rm zz}$, \ce{CH4}, \ce{CO2} and cloud profiles of the GCM inside and outside the storm regions compared to the global mean.
This allows a direct physical comparison of the properties of regions undergoing a storm phase to more quiescent regions.
As seen in Figure \ref{fig:GCM_proj}, the storm profiles are generally hotter than the background mean and this extends down to deeper pressures.
These storm regions also show similar vertical K$_{\rm zz}$ profiles, in contrast to the cold regions that show a lack of K$_{\rm zz}$. 
This suggests that the storm regions are much more convectively active than the quiet regions which gives rise to their specific chemical, cloud and temperature properties.
This is seen in the vertical profiles of \ce{CH4}, \ce{CO2} and the cloud tracer.
Here the weaker mixing in the cold regions lead to a less vertically extended and thick cloud region, which extends slightly deeper in the atmosphere due to the cooler T-p profile compared to the storm regions.
The chemical composition between the storm regions and cold parts are also different. 
The vertically quenched VMR values in the storm regions are smaller for both chemical species, suggesting generally less gas phase opacity is present in the storm regions compared to the cold areas.
Both the change in cloud structure and chemical abundance in storm areas are probably ultimately responsible for their hotter T-p profiles in general.
From these results we suggest that the formation of stormy regions also triggers a strong convective response which goes on to affect the cloud and chemical vertical profile, leading to a feedback response of heating of the atmosphere in and around the storm region.

In Figure \ref{fig:GCM_ST}, we present a Hovm{\"o}ller diagram (longitude-time sequence) of the output 42 hours past the end point of the simulation.
This shows the gradual eastward and westward propagation of the OLR and is an indication of either the zonally propagating wave patterns or the zonal jets in the atmosphere  (see a more thorough discussion in \citealp{Tan2021}).
Here we see that the overall weather structures on the planet lead to a peak in activity around the 10-30 hour mark, but then dissipate and reduce after 30 hours. 
This suggest a transition in the global wave pattern to a more quiescent phase.
This is also seen visually in the gif files\footnote{e.g. see gif files on Zenodo: \url{https://zenodo.org/records/10159911}}, where the storm regions are seen dissipating.
This also leads credence to the cause of the weakening in the spectral variability seen in Sect. \ref{sec:time_var} after hour 30.

\section{GCM synthetic observations}
\label{sec:PP}

In this section, we post-process our simulation using the 3D Monte Carlo radiative-transfer code gCMCRT \citep{Lee2022}.
We use a correlated-k approach at a spectral resolution of R = 100 between 0.2 and 30 $\mu$m.
When multiple scattering of cloud particles is included in the calculation, we use the analytical Henyey-Greenstein scattering phase function \citep{Henyey1941}.
We post-process our simulations across a 48 hour time period from the end of the simulation time, producing spectra every hour and taking into account the rotation of the planet.
We calculate spectra with cloud opacity included, as well as a set assuming no cloud opacity (but using the same 3D T-p and chemical information).
The cloud opacity is calculated assuming the same cloud size-distribution properties as simulated inside the GCM.

\subsection{Viewing angle dependence}

\begin{figure}
    \centering
    \includegraphics[width=0.49\textwidth]{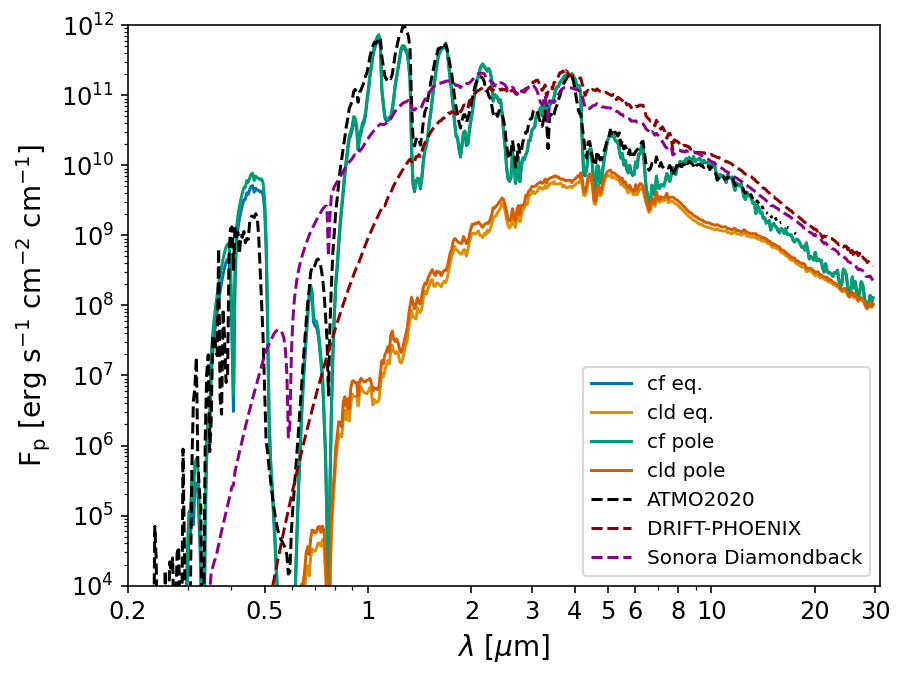}
    \caption{Planetary spectral flux from the GCM excluding cloud opacity (cf) and including cloud opacity (cld) at an equatorial viewing angle (eq.) and polar viewing angle (pole). 
    We compare to the clear atmosphere, strong mixing non-equilibrium Solar metallicity spectra from \citet{Phillips2020} (ATMO2020), the cloudy Solar metallicity spectra from \citet{Witte2009} (DRIFT-PHOENIX) and the T$_{\rm eff}$ = 1000 K, $\log$ g = 3.5, $\approx$3x Solar, f$_{\rm sed}$ = 1 Sonora Diamondback (C. Morley priv. comm.) model.}
    \label{fig:spec_1}
\end{figure}

In Figure \ref{fig:spec_1} we present spectra of our GCM simulation at the end of simulation with and without the effects of cloud opacity and at an equatorial and polar viewing angle.
Our results show that the difference in spectra between the equatorial and polar regions is only slight, with the cloud free spectra at equator and pole matching extremely well except at optical wavelengths, while a small difference in flux of maximum $\approx$10\% is seen between the polar and equatorial spectra with clouds.
The polar region tends to have a slightly higher outgoing flux compared to the equator in the cloudy case.
This is probably a direct consequence of the mid- and high-latitude regions being naturally hotter at photospheric temperatures, as well as a reduction in cloud mass at the pole (Figure \ref{fig:GCM_proj}).
Interestingly, for the cloudy case, the most difference in flux between the equator and pole occurs at the silicate absorption band at $\sim$10$\mu$m.

In Figure \ref{fig:spec_1}, we compare our model results without cloud opacity to the cloud free ATMO2020 grid Solar metallicity \citep{Phillips2020} spectra at T$_{\rm eff}$ = 1000K, log g = 3. 
Here the effect of increased metallicity on the atmosphere is apparent as the T-p structures are similar between the models (Figure \ref{fig:GCM_1D}), but deeper molecular spectral features and alkali lines occur in the GCM simulation, with a generally reduced overall outgoing flux across the wavelength range.

Figure \ref{fig:spec_1} also shows the spectra from the DRIFT-PHOENIX set of models \citep{Witte2009} that includes the microphysical cloud model and opacity of \citet{Helling2008}.
The overall shape of the spectra is similar, suggesting that the reddening effect of the cloud is similar between the models, but the overall flux is reduced by an order of magnitude or more in the GCM. 
This suggests the overall cloud opacity for this regime is much less in the DRIFT-PHOENIX model, possibly as a result of the increased metallicity used in the GCM simulation providing a lager condensable cloud mass.
Comparing the to Sonora Diamondback model (C. Morley priv. comm.), where we chose a vertically well mixed cloud structure with f$_{\rm sed}$ = 1 \citep{Ackerman2001,Rooney2022} most like the well mixed cloud structure from the GCM.
The Sonora Diamondback model has a hotter T-p profile (Figure \ref{fig:GCM_1D}), which pushes the spectra towards bluer wavelengths compared to the GCM results, however, the overall shape and relative strength of the spectral features to the continuum in this cloudy model are comparable to the GCM results.
In addition, both the DRIFT-PHOENIX and Sonora Diamondback spectra do not show strong infrared silicate absorption features unlike the GCM results.

\subsection{Mini-chem vs CE}

\begin{figure}
    \centering
    \includegraphics[width=0.49\textwidth]{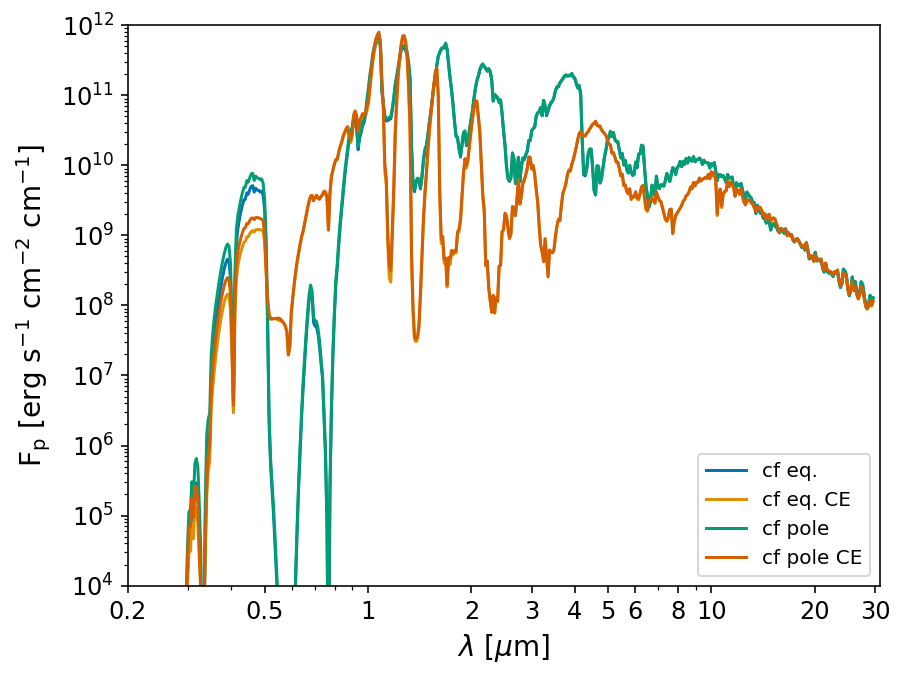}
    \caption{Cloud opacity free (cf) spectral post-processing of the GCM simulation at the equator (eq.) and pole (pole), assuming the mini-chem results and chemical equilibrium (CE).
    The main differences between the chemical non-equilibrium and equilibrium spectra occur in the \ce{CH4} and \ce{NH3} spectral features, as well as Na and K.}
    \label{fig:CE}
\end{figure}

In Figure \ref{fig:CE} we compare spectra of the GCM output without cloud opacity to the GCM output assuming chemical equilibrium. 
We use the \textsc{FastChem} chemical equilibrium code \citep{Stock2022,Kitzmann2023} to calculate the chemical species volume mixing ratio from the GCM output.

The difference of the spectra between the chemical non-equilibrium and equilibrium model is stark, mostly stemming from the lack of \ce{CH4} and \ce{NH3} absorption features in the mini-chem results, brought on by the vertical quenching of these species (Figure \ref{fig:GCM_1D}).
\ce{CO2} absorption is also present in the mini-chem results, not seen in the equilibrium calculation.
The Na and K features are also much weaker in the equilibrium results.
\ce{H2O} features are also stronger in the equilibrium case, as quenching reduces the mixing ratio of \ce{H2O} in the upper atmosphere (Figure \ref{fig:GCM_1D}).

\subsection{Time variability}
\label{sec:time_var}

\begin{figure*}
    \centering
    \includegraphics[width=0.49\textwidth]{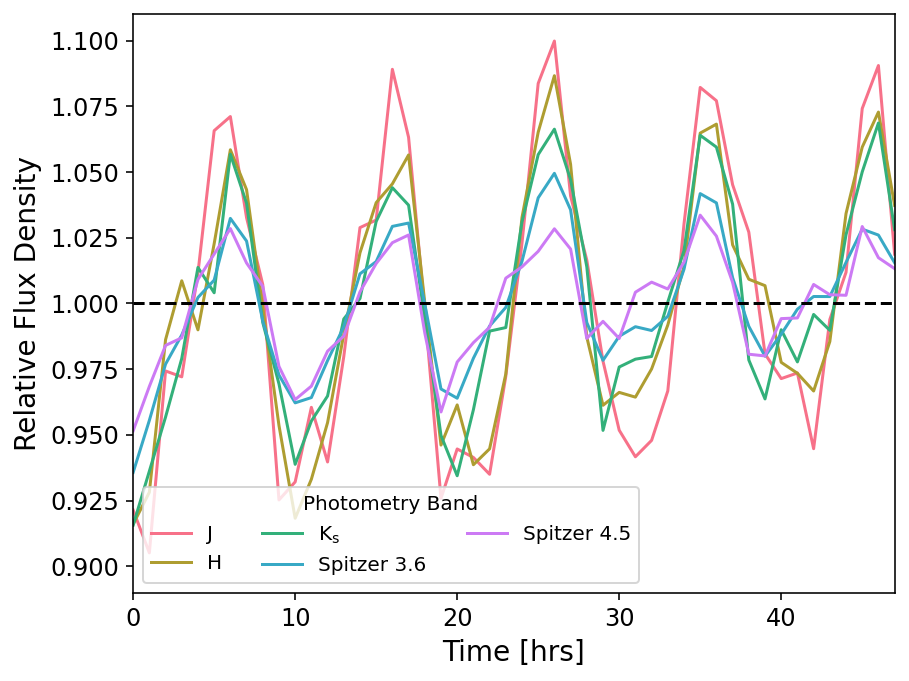}
    \includegraphics[width=0.49\textwidth]{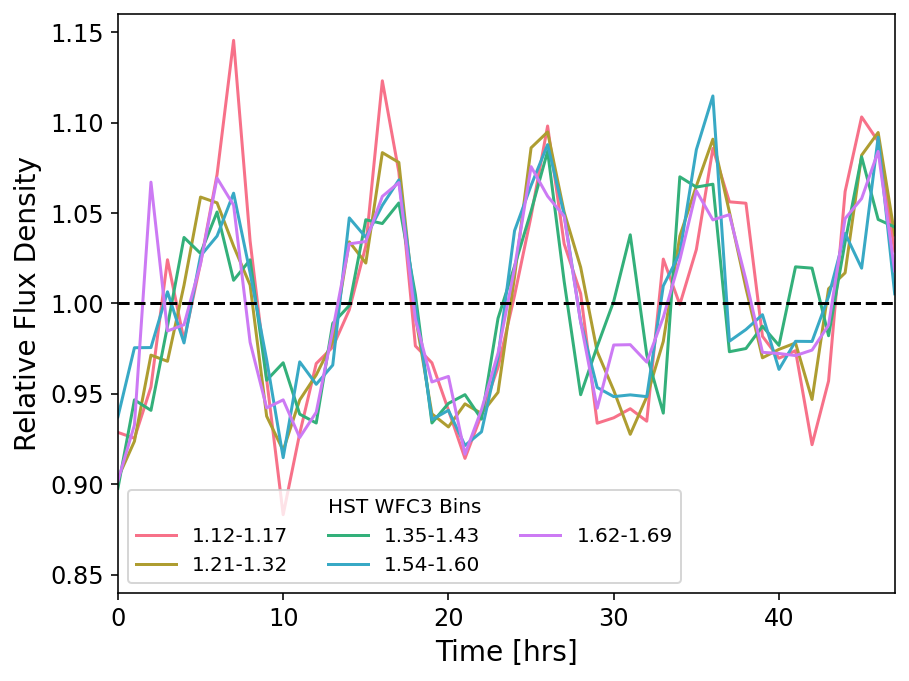}
    \includegraphics[width=0.49\textwidth]{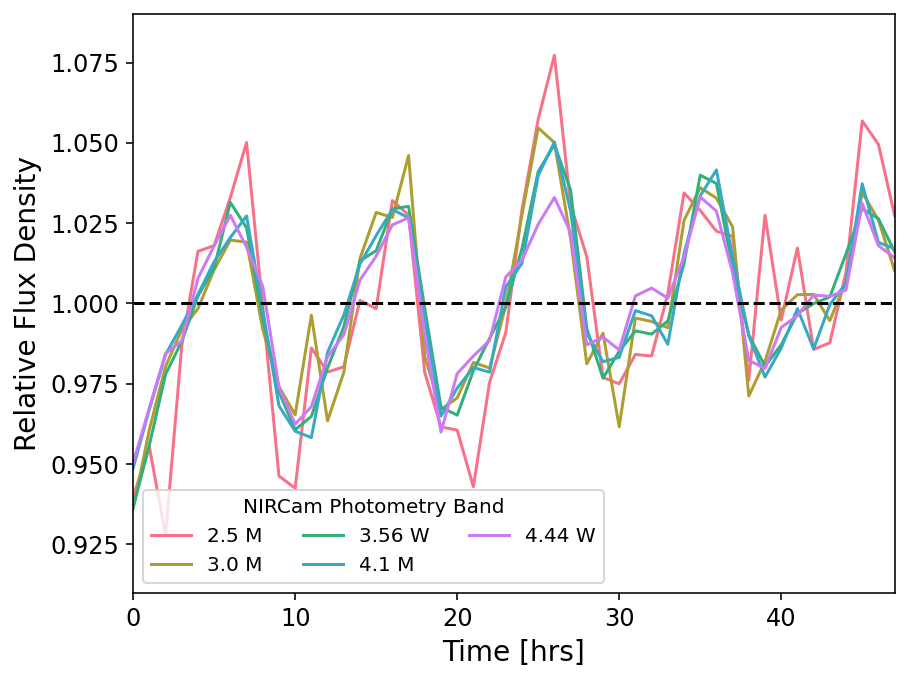}
    \includegraphics[width=0.49\textwidth]{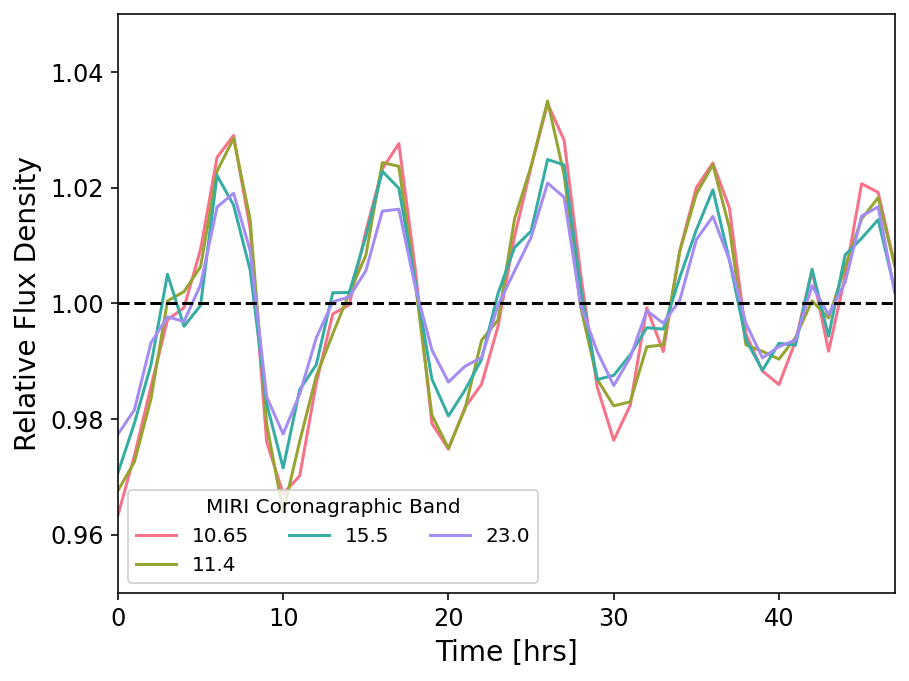}
    \caption{Normalised relative flux density against time in hours resulting from the GCM simulation for an equatorial viewing angle.
    Top left: J, H, K$_{\rm s}$ and Spitzer 3.6 and 4.5 photometric bands.
    Top right: HST WFC3 wavelength bins.
    Bottom left: Selection of NIRCam bands, the same as \citet{Carter2023}.
    Bottom right: All MIRI coronagraphic mode bands.}
    \label{fig:time_vari}
\end{figure*}

In this section, we present the time-variability properties of the GCM simulation by calculating normalised relative flux density values with time for several photometric bands used for variability studies in the literature.
We chart the spectral band fluxes taking into account the rotation of the planet and evolution of weather across a 48 hour span of time.
In \cite{Tan2021} it was found that an equatorial viewing angle produced the highest signals of variability.
We therefore post-process at an equatorial viewing angle and include cloud opacity and multiple-scattering when calculating the flux to try produce the largest variability in the spectra.

Figure \ref{fig:time_vari} shows the normalised flux density with time of our simulation, with each plot showing a different selection of photometric bands commonly used in the literature, from the ground based J, H, K$_{\rm s}$ used in surveys \citep[e.g.][]{Radigan2014} to the commonly used HST WFC3 bins \citep[e.g.][]{Buenzli2012} and the Spitzer 3.6 and 4.5 photometric bands.
In addition, Figure \ref{fig:time_vari} contains predictions for JWST instruments, where we chose the same NIRCam bands as \citet{Carter2023} and all MIRI coronagraphic bands.

Overall, it is clear that different bands show different peak-to-peak fractional variability as seen in the observational data of brown dwarf objects \citep[e.g.][]{Apai2017, Zhou2022}.
Optical and near-IR bands show the most relative variability at maximum $\sim$10\%.
The J band shows the largest variation, at $\sim$15\% peak-to-trough, comparable to some of the largest observed variability at near-IR wavelengths \citep[e.g.][]{Apai2017}.
In the mid-IR, the Spitzer and MIRI bands show around a $\sim$4-5\% variability for all bands, with changes in the amplitude with time.
These bands also show less scatter than the optical and near-IR bands, suggesting they probe more quiescent regions of the planet.
A trend of decreasing amplitude with time is also most seen in the mid-IR bands, suggesting the atmosphere is in a dissipating storm phase of evolution, relaxing back to a background mean value.
This is confirmed through visual analysis of the time-dependent GCM output\footnote{e.g. see gif files on Zenodo: \url{https://zenodo.org/records/10159911}}.
Our spectra also show signs of phase differences between bands, suggesting the weather and rotational features affect the wavelength dependence of the planet at different response rates.
Most obviously, each band shows the rotational component of the variability well, with a period of 10 hours from crest-to-crest or trough-to-trough, the same as the GCM rotational rate.

\section{Discussion}
\label{sec:disc}

In this initial single planet demonstration, through fully coupling time-dependent chemistry, cloud and radiative-transfer feedback, our current model represents some of the most self-consistent 3D exoplanet atmosphere simulations to date.
Lastly, by using the gCMCRT 3D spectral radiative transfer model \citep{Lee2022}, we are able to consistently post-processes the GCM results to produce spectral and photometric band weighted light curve predictions of the exoplanet.
This improves on the broadband light curves of previous theoretical studies which can not be directly compared to observational data.
By building a library of spectral models in the future, band variability predictions like the ones produced in this paper will be valuable for legacy and JWST data sets and enable further characterisation of brown dwarf and exoplanet atmospheric dynamics.

Overall, our results suggest a strong connection and feedback between the spatial cloud coverage and chemical composition of the atmosphere, in particular where storm regions form and dissipate at mid-latitude regions.
We propose two mechanisms that explain the feedback between the cloud particles and chemical composition:

Firstly, it is clear that the cloud particles and chemical species are both dynamically and temperature driven.
In hot Jupiter models the strong day-night winds homogenise the chemical composition across the globe \citep{Drummond2020,Zamyatina2023,Lee2023a,Tsai2023} in both the horizontal and meridional directions.
The relevant timescales for the hot Jupiter case is the balance between the advective timescale and chemical timescale, where if the advective timescale is short compared to the chemical timescale, chemical homogenisation can be expected.
\citet{Zamyatina2023} provide evidence for this case, where a `sweet spot' (at around the HD 189733b parameter range) for non-equilibrium chemical behaviour was found in their hot Jupiter sample, planets hotter and cooler than this range showed less non-equilibrium behaviour in the horizontal direction.

A similar effect occurs here, with the convective response in storm regions allowing the atmosphere to vertically drive the chemical non-equilibrium quenching behaviour, leading to chemical differences in storm regions compared to quiescent regions.
This effect appears to be most strong when large storms form and drives a maximal cloud coverage, and therefore largest cloud opacity.
Chemical species have to respond to the change in mixing environment of the local atmosphere in the storm region, this can lead to a changes in the quench level and the quenched VMR value of each species in these regions.
Once the storms dissipate however, the chemical species have time to relax and return to the background mean values.
This chemical timescale storm behaviour was also seen in LLT, but at a much lower intensity compared to the current model.
Overall, our simulation suggests that the increased intensity and convective strength of the storms due to the presence of cloud increases the non-equilibrium behaviour and inhomogeneity of the chemistry in the atmosphere.

Secondarily, the clouds have a direct effect on the local temperature in the atmosphere, which in turn enables the strong convective response in a time-dependent fashion, as seen in even 1D models \citep{Tan2019}.
The cloud opacity invokes temperature fluctuations associated with the storm formation, increasing the vertical profile temperature in the storm regions.
The chemical network must respond to this change in temperature, also contributing to the differences in species VMR between the storm and quiescent regions.
We suggest that the increased cloud opacity in stormy regions also alters the radiative response of the atmosphere, allowing the storm regions to survive longer and be more vigorous here compared to LTT.
Conversely, should a colder patch form with less thick cloud coverage, the network will also respond in kind.
Overall, this leads to the inhomogeneous behaviour of the chemical species across the globe as species react to changes in temperature profile and mixing rates through the mechanism of storm formation.

The rotation rate of the exoplanet or brown dwarf plays a key role in the dynamical properties and variability properties of the atmosphere \citep{Tan2021b,Tan2022}.
\citet{Hammond2023} perform shallow water models to explore the dynamical regimes of isolated brown dwarfs and exoplanets.
They show four distinct dynamical regimes as a function of the thermal Rossby number and radiative-timescale.
Our simulation falls in their first regime of large vortices and banded jets at the dynamical layers.
Performing our GCMs across this parameter space will be an important future consideration to fully explore the cloud and chemistry connections across each dynamical regime.

In this study, we have included time-dependent mixing length theory directly into the GCM, which replaces the dry convective adjustment scheme from LTT.
The MLT derived eddy diffusion coefficient, K$_{\rm zz}$, was then used to estimate the vertical mixing of gas phase tracers due to small scale convective motions, not captured by the base GCM simulation itself.
We suggest this framework could be utilised by other cloud models in use for 3D exoplanet models such as EddySed \citep[e.g.][]{Lines2019,Christie2023} and mini-cloud \citep{Lee2023c}.
3D mixing studies such as those in \citet{Parmentier2013}, can also be updated to include the component, moving beyond mixing estimations that only use vertical velocities from the GCM.
However, we note MLT is a localised process rather than fully vertically coherent, and approximates an advective process (convection) with a diffusive approach \citep{Joyce2023}. 
Our simple 1D MLT formalism used in this study also neglects important 3D effects such as convective shear and rotational effects \citep[e.g.][]{Prat2014} and convective plume patchiness effects \citep{Lefevre2022}.
With various different formalisms also available for MLT \citep{Joyce2023}, it may be important to find calibrated mixing length parameters for low-gravity exoplanets, similar to what was done for stars using convection resolving models \citep[e.g.][]{Ludwig1999}.
The convective overshoot parameters may also be different for low-gravity objects.
In addition, it may also be important to take into account chemical gradient effects that can trigger small scale convection \citep{Prat2014,Tremblin2015}.

Despite the fullness of our simulation, several considerations for the cloud model are currently not included.
For our T-dwarf regime other cloud species such as MnS, \ce{Na2S}, Cr and KCl may also be present as their saturation curves pass through our simulation T-p profiles.
The results of \citet{Morley2012} suggest that these clouds can be important in setting the observed spectra and T-p profiles in this regime.
Future iterations can extend our simple equilibrium cloud scheme to include multiple sets of vapour and cloud components.
However, we note that microphysical studies show the size-distribution properties of each species are different \citep{Powell2018}, this complicates the process of adding multiple species to the scheme in a consistent manner.
Another option is to include a microphysical cloud model such as mini-cloud \citep{Lee2023c} or CARMA \citep{Gao2018}, but this is likely to increase the computational expense of the model significantly.
The results of \citet{Woitke2020} also suggest a long convergence time of a few simulated years required for such models.
If these more volatile cloud species form at high altitude, then they may affect the variability of the atmosphere more than simulated here, especially at mid-infrared wavelengths that probe the cooler temperatures these clouds would form at.

Our resulting spectra show a highly reddened object and large reduction in the outgoing flux when cloud opacity is included, not seen when comparing to the 1D RCE DRIFT-PHEONIX and Sonora Diamondback models that contain a cloud component.
We suggest this strong cloud opacity effect seen in our model is due to two components:
\begin{itemize}
    \item The 10x Solar metallicity assumed in the simulation, allowing a larger cloud mass to condense.
    \item The specific log-normal size-distribution parameters chosen in this study, where the assumption of a 1$\mu$m effective particle size resulted in a settling rate that could not compete with the strong vertical mixing rate in the GCM.
\end{itemize}
This combination resulted in a thick, well vertically mixed cloud layer, whereas if we assumed a larger effective particle size, a more compact cloud layer at higher pressures would form, resulting in a less red spectra.
However, despite this, significant variability was seen in all photometric bands, suggesting even thickly cloud covered and highly reddened exoplanets show detectable tell-tale signs of variability in their atmospheres.
This initial model only shows one specific set of cloud parameters, and future studies using this framework can apply different parameters to see their effects using the lessons learned from this study.
We note that directly imaged and field objects commonly show extremely red spectral colours \citep[e.g.][]{Schneider2023}, suggesting that our simulated spectra are not outside of possibility for these objects.

Future developments of this model framework can attempt to address some of these shortcomings, as well as perform simulations across a multitude of planetary parameters.
In particular, we envision future band variability predictions from these types of GCM to be a valuable comparison to observational data, able to provide a useful characterisation method to attempt to discover which weather patterns and dynamical regimes of the GCM fit the data best.

\section{Summary \& Conclusions}
\label{sec:conc}

In this follow up to \citet{Lee2023a}, we improve the realism of isolated and companion brown dwarf and exoplanet GCM simulations through the inclusion of several processes; mixing length theory, convective mixing of tracers through a diffusive scheme, a tracer based equilibrium cloud model and kinetic chemistry.
These processes are consistently coupled to a non-grey, spectral radiative-transfer scheme able to provide radiative-feedback onto the atmosphere from the cloud and non-equilibrium chemistry scheme.
With this combination of time-dependent physical processes, this study represents one of the most self-consistent and complete contemporary 3D models of exoplanet atmospheres, on par with the sophistication and included physics of 1D radiative-convective models in use in the field, for example, equilibrium cloud modelling \citep[e.g.][]{Brock2021}, kinetic chemistry \citep[e.g.][]{Drummond2016} and radiative-feedback \citep[e.g.][]{Mukherjee2023}.

To exhibit the new additions to the model, we simulate a T$_{\rm int}$ = 1000 K, metal enhanced young hot sub-Jupiter companion object. 
We find a complex feedback between clouds, chemistry and the temperature structure of the atmosphere. 
Global averages of the simulated atmosphere align well with current 1D RCE modelling at a similar regime, but the time-dependent properties of the GCM reveal a highly inhomogenous and stormy atmosphere, with the clouds and chemistry responding to changes in the local thermochemical environments brought on by the formation and dissipation of stormy mid-latitude regions.

We post-process our GCM output using the 3D gCMCRT model and produce photometric band variability curves.
We find the rotational and weather patterns in the GCM affect the peak-to-peak variability of the bands significantly, with up to $\sim$20\% seen in optical and near-IR bands and $\sim$5\% or less in mid-IR bands.
Our model shows spectral phase lag behaviour as well as differences in the variability amplitude between bands as noted by several observational campaigns of brown dwarf objects.

Our model set-up can be easily incorporated for modelling upcoming JWST targets of isolated/companion brown dwarf and exoplanet atmospheres across the parameter regime.
However, we note for highly irradiated planets an important missing piece of physics from the current model is the inclusion of photochemical processes, which is known to have important global chemical impacts when coupled to flows \citep[e.g.][]{Baeyens2022, Tsai2023b}.

Overall, the prospect of discovering and characterising companion young sub-Jupiter exoplanets is a highly anticipated capability of JWST.
Our updated, more complete, modelling approach provides a useful insight into the complex, coupled feedback between the atmospheric dynamics, cloud and kinetic chemistry processes occurring in these atmospheres, showing these worlds will be an important part of a full understanding of exoplanet atmospheres across their diverse astrophysical environments.

\section*{Acknowledgements}
E.K.H. Lee is supported by the SNSF Ambizione Fellowship grant (\#193448).
S-M. Tsai is supported by the University of California at Riverside.
We thank D. Christie for extensive testing of mini-chem.
We thank C. Morley and team for providing a selection of Sonora Diamondback models in advance of their publication.
We thank M. Steinrueck and M. Lef\`{e}vre for comments and advice on the project and manuscript.
The HPC support staff at the University of Bern is highly acknowledged.
Calculations were performed on UBELIX (\url{http://www.id.unibe.ch/hpc}), the HPC cluster at the University of Bern.

\section*{Data Availability}
The 1D radiative-transfer, mini-chem, gCMCRT and various other source codes are available on the lead author's GitHub: \url{https://github.com/ELeeAstro}.
Gif and simulation output in NetCDF format is available from Zenodo: \url{https://zenodo.org/records/10159911}.
All other data is available upon request to the lead author.


\bibliographystyle{mnras}
\bibliography{bib} 





\bsp	
\label{lastpage}
\end{document}